\documentclass[prb,aps,final,twocolumn,nofootinbib,superscriptaddress]{revtex4-1}
\usepackage{amsmath}
\usepackage{natbib}
\usepackage[pdftex]{graphicx}
\usepackage{epstopdf}
\usepackage{times}
\usepackage{txfonts}
\usepackage{xcolor}
\usepackage{textcomp}
\usepackage[colorlinks, bookmarks=false, citecolor=blue, linkcolor=red,urlcolor=blue]{hyperref}

\def\be {\begin{equation}}
\def\ee {\end{equation}}

\def\horparallel{ \lower.5ex\hbox{ \includegraphics[width=2ex]{fig-hor.pdf}}\,\,
}
\def\vertparallel{ \lower.5ex\hbox{
\includegraphics[width=2ex]{fig-vert.pdf}}\,\, }
\def\gsim{\mathrel{\rlap{\lower4pt\hbox{\hskip1pt$\sim$}}\raise1pt\hbox{$>$}}}
\def\lsim{\mathrel{\rlap{\lower4pt\hbox{\hskip1pt$\sim$}}\raise1pt\hbox{$<$}}}

\newcommand{\dagga}{{\phantom{\dagger}}}

\begin{document}

\title{Semiclassical evidence of columnar order in the fully frustrated transverse field Ising model on the square lattice}

\author{Tommaso Coletta}
\affiliation{Institute of
  Theoretical Physics, \'Ecole Polytechnique F\'ed\'erale de Lausanne
  (EPFL), CH-1015 Lausanne, Switzerland}
\author{Sergey~E. Korshunov}
\affiliation{L.~D. Landau Institute for Theoretical Physics RAS, 142432
Chernogolovka, Russia}
\author{Fr\'ed\'eric Mila}
\date{\today}
\affiliation{Institute of
  Theoretical Physics, \'Ecole Polytechnique F\'ed\'erale de Lausanne
  (EPFL), CH-1015 Lausanne, Switzerland}


\begin{abstract}
 We investigate the zero-temperature phase diagram of the fully frustrated transverse field Ising model on the square lattice both in the classical limit and in the presence of quantum fluctuations.
 At the classical level (the limit of infinite spin $S$), we find that upon decreasing the transverse field $\Gamma$ this model exhibits a phase transition from the fully polarized state into an eight-fold degenerate translational symmetry breaking state.
 This phase can be identified to correspond to plaquette order in the dimer language and remains the lowest-energy state in the entire range of fields below the critical one, $\Gamma_c$.
 The eight-fold degenerate solution which corresponds to columnar order in the dimer language is a saddle point of the classical energy.
 It is degenerate with the plaquette solution at $\Gamma=0$ and is only slightly higher in energy in the whole interval $0<\Gamma<\Gamma_c$.
 The effect of quantum fluctuations is investigated in the context of a large S expansion both for the plaquette and columnar structures.
 For this purpose we employ an approximate method allowing to estimate from above the fluctuation-induced correction to the energy of a configuration which at the classical level is a saddle point of the energy, \textit{not} a local minimum, and find that the harmonic quantum fluctuations show
 a clear tendency to overcome the energy difference between the two states. 
 For relatively high fields the transition from the plaquette to the columnar state takes place at values of $S$ large enough to be in the domain of validity of 
 the harmonic approximation. 
\end{abstract}

\maketitle

\section{Introduction}
Fully frustrated transverse field Ising models \cite{MoessnerSondhiChandra00,MoessnerSondhi01} (FFTFIMs) are characterized by the interplay of two ingredients: a frustrated Ising-like interaction between
neighboring spins 1/2 and a field term coupled to the transverse component of the spin which is the source of quantum dynamics in the system.
In magnetic systems, the term {\em fully frustrated} implies that there is an odd number of antiferromagnetic (AF) bonds on each elementary plaquette, which makes it impossible to satisfy simultaneously all bonds on any of them.
Such a model  was first introduced by Villain under the name of `odd model'.\cite{Villain77}
However, the investigation of fully frustrated Ising models has started long before that: the exact solution of the classical AF Ising model on the triangular lattice (evidently belonging to this class) has been constructed already in 1950.\cite{Wannier50,Houtappel50}

Starting from a classical Ising model, one can obtain its quantum version
by including a transverse field term. Thus, the \makebox{FFTFIM} is described by the Hamiltonian
\begin{equation}\label{eq:Quantum Hamiltonian}
 \mathcal{H}=\sum_{\langle i,j\rangle}{J_{ij}\sigma_i^z\sigma_j^z}-\Gamma\sum_i \sigma_i^x,
\end{equation}
where $\sigma_i$ are spin $1/2$ operators defined on sites $i$ of some regular lattice and the first sum runs over the pairs of nearest neighbors on this lattice.
The magnitude of the coupling $|J_{ij}|=J$ is the same on all bonds of the lattice while its sign is positive for an odd number of bonds of each plaquette and negative for the remaining bonds.
All Ising models on the same lattice satisfying this rule can be transformed into each other by a gauge transformation and accordingly are equivalent. \cite{Villain77} This leaves one with the freedom to choose the most convenient gauge, that is, which bonds (satisfying the ``odd rule``) to consider as antiferromagnetic.
References \onlinecite{MoessnerSondhiChandra00} and \onlinecite{MoessnerSondhi01} review the properties of FFTFIM on a number of periodic two-dimensional lattices.

One of the main interests in FFTFIMs comes from the fact that they are closely related to the quantum dimer model (QDM)
of Rokhsar and Kivelson\cite{RokhsarKivelson88} first introduced in the context of high $T_c$ cuprate superconductors.
Moessner, Sondhi and Chandra \cite{MoessnerSondhiChandra00} showed that in the limit $\Gamma/J\rightarrow0$ FFTFIMs can
be mapped on purely kinetic QDMs (i.e. QDMs whose potential energy $V$ is equal to zero) on the dual lattice. A detailed review
of the mapping between the FFTFIM on the honeycomb lattice and the purely kinetic QDM on the triangular lattice can be found in Ref.~\onlinecite{Misguich08}.
In contrast to the triangular-lattice QDM, for which the ground state symmetry at $V=0$ is well known, \cite{Moessner01,Ralko05,*Ralko06,*Ralko07}
the nature of the ground state in the purely kinetic square-lattice QDM has long remained controversial, with different numerical studies predicting columnar,\cite{Sachdev89,Syljuasen06} plaquette\cite{Leung96} or mixed type\cite{Ralko08} order.

In the present work, the competition between the plaquette and columnar orders in the FFTFIM on the square lattice is systematically investigated in the context of a large $S$ expansion.
The main results of our work were briefly announced earlier in Ref.~\onlinecite{Wenzel12} (by Sandro Wenzel and the authors of the present article) focused mainly on large scale quantum Monte Carlo simulations of the FFTFIM on the square lattice.

Reference \onlinecite{Wenzel12} argues in favor of the existence of only two phases in the zero-temperature phase diagram of this model as the ratio $\Gamma/J$ is varied.
For low fields, a state which in terms of the closely related quantum dimer model has columnar character is stabilized.
\footnote{A similar conclusion was reached earlier by Jalabert and Sachdev \cite{Sachdev91} for the three-dimensional classical
model (constructed by stacking the fully frustrated Ising model) which, at finite temperatures, can be expected to behave as the
zero-temperature quantum FFTFIM. However, as pointed out in Ref.~\onlinecite{Wenzel12}, the numerical analysis of Ref.~\onlinecite{Sachdev91} did not resolve the phase of the order parameter used in this work, without which it is impossible to distinguish the columnar, plaquette and mixed phases from each other.}
As the field is increased beyond a critical value, the ordering disappears and the system becomes uniformly polarized.

The present work confirms the conclusion on the columnar nature of the ordering in the FFTFIM on the square lattice by showing that,
to first order in $1/S$, quantum fluctuations have the general tendency to stabilize columnar order. This conclusion has to be taken as only
indicative for weak transverse field because of strong quantum fluctuations and anharmonic contributions, but for not too weak field it has more solid grounds
and therefore can be considered as an analytical confirmation of the results of Ref.~\onlinecite{Wenzel12} .

The paper is organized as follows. Sec.~\ref{sec:Classical Limit} discusses the classical version of the model corresponding to the limit of infinite spin $S$.
The polarized state with all classical spins aligned along the field extends down to $\Gamma_c=2\sqrt{2}J$.
Our analytical analysis both in the critical region just below $\Gamma_c$ and in the low field limit, $\Gamma/J\ll1$, demonstrates that the classical
ground state corresponds to a plaquette structure in the dimer language. The results of the numerical energy minimization for a single plaquette are then used to prove that the same conclusion is valid in the whole interval $0<\Gamma<\Gamma_c$. On the other hand, the spin configurations corresponding to columnar order in the dimer language are identified to be saddle points of the classical energy. 
In the vicinity of $\Gamma_c$, the degeneracy between the columnar and plaquette states is lifted only when the expansion of the classical energy is extended up to the eighth order, the difference in energies scaling like $(\Gamma_c-\Gamma)^4/J^3$.

Sec.~\ref{sec:Semiclassical approach} studies the effect of quantum fluctuations around the plaquette and columnar structures. Harmonic fluctuations in the plaquette state are treated via the standard linear spin-wave approximation.
The same approach is inapplicable for analyzing the fluctuations in the columnar state as it is a saddle point of the classical energy, not a local minimum.
This forces us to restrict ourselves with finding an upper estimate for the zero-point energy of the state with columnar structure. Within the framework of this approximation, the harmonic fluctuations turn out to stabilize the columnar state over the plaquette one in the entire field range for $S=1/2$.
The amplitude of harmonic fluctuations as well as the relevance for the QDM are also discussed in this section.
A short conclusion is given in Sec. \ref{sec:Conclusion}.

\section{Classical limit}\label{sec:Classical Limit}

In this section we discuss the classical ground state of the model for
different values of the ratio $\Gamma/J$. Throughout this paper by {\em classical version} of the model we mean not the classical fully frustrated Ising model one obtains when putting $\Gamma=0$, but the model whose Hamiltonian can be obtained by replacing in Eq. (\ref{eq:Quantum Hamiltonian})
the Pauli matrices by the corresponding components of vectors of unit lengths. Note that at the conceptual level this simple transformation can be interpreted as consisting of two separate steps. Firstly, the quantum \makebox{FFTFIM} discussed in the introduction is generalized to the case of spins of arbitrary
magnitude $S$ by rewriting Eq. (\ref{eq:Quantum Hamiltonian}) as
\begin{equation}\label{eq:Generalized Quantum Hamiltonian}
 \mathcal{H}=\frac{1}{S^2}\sum_{\langle
i,j\rangle}{J_{ij}S_i^zS_j^z}-\frac{\Gamma}{S}\sum_i S_i^x\;,
\end{equation}
secondly one takes the classical limit,
\makebox{$S\rightarrow\infty$}, in which the fluctuations are suppressed. In this limit, the normalized spin
operators ${\bf S}_i/S$ are replaced by three-dimensional vectors ${\bf
n}_i$ of unit length. Below we refer to
these variables as classical spins. In terms of classical spins the
original Hamiltonian reduces to
\begin{equation}\label{eq:Generalized Classical Hamiltonian}
 \mathcal{H}=\sum_{\langle i,j\rangle}{J_{ij}n_i^z n_j^z}-\Gamma\sum_i
 n_i^x\;
\end{equation}
which has exactly the same structure as (\ref{eq:Quantum Hamiltonian}),
but with the spin operators replaced by classical spins ${\bf n}_i$ with $|{\bf n}_i|=1$.

In contrast to the analogous model on the triangular lattice, the fully
frustrated Ising model on the square lattice does not have any ``natural"
gauge, hence it is convenient to discuss the structure of the ordered phases in this
model in terms of some gauge-invariant variables. For this purpose, following Refs.
\onlinecite{Misguich08},\onlinecite{Wenzel12} and \onlinecite{Coletta11}, we introduce the quantity
\begin{equation}\label{eq:Dimer Density}
 d_{ij}=\frac{1}{2}\left(1+\frac{J_{ij}}{J}n_i^zn_j^z\right)\in[0,1]\;,
\end{equation}
which is equal to 0 if the energy of the bond $\langle i,j\rangle$ is
negative (that is, is equal to $-J$) and to 1 if this bond is frustrated,
that is, has positive energy $+J$. In terms of the quantum dimer model
\cite{RokhsarKivelson88} to which the spin-1/2 quantum Ising model (\ref{eq:Quantum
Hamiltonian}) is equivalent \cite{MoessnerSondhiChandra00} in the limit $\Gamma/J \to 0$,
the variables $d_{ij}$ defined in Eq. (\ref{eq:Dimer Density}) correspond
to the occupation number of dimers on the bond of the dual lattice (on
which the dimer model is defined) crossing the bond $\langle i,j\rangle$
on the Ising model lattice. For this reason, the variables $d_{ij}$ can be
called dimer densities.

In the remaining part of this section we analyze the structure of the
ground state of the classical model (\ref{eq:Generalized Classical
Hamiltonian}) on the square lattice for different values of the ratio $\Gamma/J$. When it is necessary to specify a particular gauge, we always use the simplest periodic gauge in which the antiferromagnetic bonds occupy every second vertical line of the lattice whereas all other bonds are ferromagnetic.
This pattern of couplings is depicted in Fig.~\ref{fig:Unit cells} where
dashed lines represent $J_{ij}=+J$ and solid lines $J_{ij}=-J$. In the
chosen gauge the lattice is split into two nonequivalent sublattices and
accordingly its unit cell has a rectangular shape and includes two
sites, see Fig.~\ref{fig:Unit cells}(a).

\subsection{Large $\Gamma/J$ case and critical modes}\label{sec:Ginzburg-Landau}
In this section we discuss the structure of the classical ground state of
the model (\ref{eq:Generalized Classical Hamiltonian}) for large values of
the ratio $\Gamma/J$.
Since for any $\Gamma>0$, the energy of any spin with $n_i^y\neq 0$ can be
decreased by rotating it around the $z$ axis until $n_i^y$ vanishes, in the
lowest energy states all spins must be oriented in the $xz$
plane. Accordingly, in all configurations that we have to compare, the
classical spins can be parametrized by their projections on the $z$ axis,
\begin{equation}\label{eq:parametrization}
{\bf n}_i=\left(\sqrt{1-(n_i^z)^2},\;0\;n_i^z\right)
\end{equation}
which allows one to rewrite the expression for the classical energy given by
Eq. (\ref{eq:Generalized Classical Hamiltonian}) as
\begin{equation}\label{eq:E classical}
 E=\sum_{\langle i,j\rangle}{J_{ij}n_i^z n_j^z}-\Gamma
\sum_i{\sqrt{1-(n_i^z)^2}}\,.
\end{equation}

\begin{figure}
 \centering
\includegraphics[width=7cm]{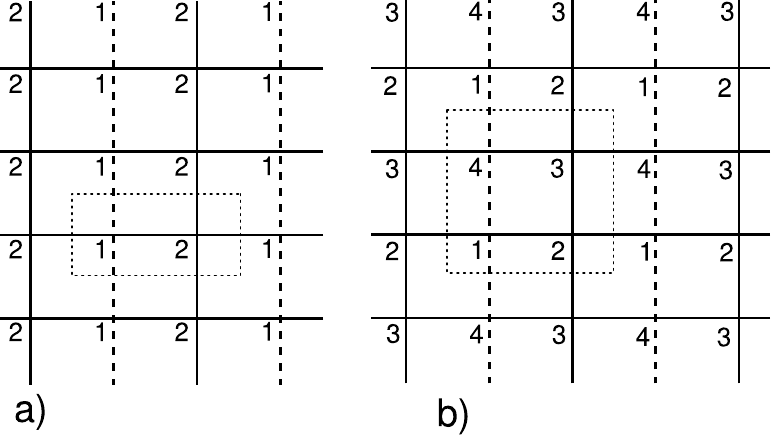}
\caption{(Color online) Two (a) and four (b) site unit cells used in calculations for the choice
of gauge used in this work (AF bonds, $J_{ij}=J$, are depicted by dashed
lines and F bonds , $J_{ij}=-J$ by solid lines).
}
\label{fig:Unit cells}
\end{figure}

It is evident that in the infinite transverse field limit the minimum of the
energy is achieved when $n_i^z=0$. However, upon decreasing $\Gamma/J$,
the spins can acquire a small component in the $z$ direction. Expanding
the square roots in Eq. (\ref{eq:E classical}) in powers of $n_i^z$ yields
\begin{equation}\label{eq:Energy to 8th order real space}
\begin{array}{lll}
  E&=&\displaystyle E_\textrm{pol}+\sum_{\langle i,j\rangle}{J_{ij}n_i^z n_j^z} \\
   & &\displaystyle +\Gamma\sum_i\left[{\frac{{(n_i^z)}^2}{2}+\frac{{(n_i^z)}^4}{8}+\frac{{(n_i^z)}^6}{16}+\frac{5{(n_i^z)}^8}{128}}+\ldots\right]\;,
\end{array}
\end{equation}
where $E_\textrm{pol}=-\Gamma N$ is the energy of the fully polarized
state, $N$ being the total number of lattice sites. Introducing the
Fourier transforms of $n_i^z$ via $n_{i}^z=\sum_{{\bf q}}n_{{\bf
q},m}^ze^{i{\bf q}{\bf r}_i}$ [with index $m=1,2$ labeling the
sublattice to which site $i$ belongs, see Fig.~\ref{fig:Unit cells}(a)], the energy per site
$\mathcal{E}$ of the system with periodic boundary conditions can be
rewritten as
\begin{equation}\label{eq:Energy to 8th order}
\begin{array}{lll}
 \mathcal{E}&=&\displaystyle -\Gamma -\frac{J}{4}\sum_{{\bf q},m,m^\prime}{n_{-{\bf q},m}^z
\left[M({\bf q})-\frac{\Gamma}{J}\right]_{m,m^\prime} n_{{\bf q},m^\prime}^z} \\
& &
+\displaystyle \frac{\Gamma}{16}T_4 + \frac{\Gamma}{32} T_6 + \frac{5\Gamma}{256} T_8 +\ldots
\end{array}
\end{equation}
where $T_{\mu}$ denotes the sum
\begin{equation}\label{eq:Tmu}
 T_{\mu}=\sum_{m=1}^2{\prod_{\nu=1}^{\mu}{\left(\sum_{{\bf q}_\nu}{n_{{\bf q}_\nu,m}^z}\right)}} \cdot \sum_{\bf G} \delta_{\sum_{\tau=1}^\mu{{\bf q}_\tau},{\bf G}} ~,
\end{equation}
where the last sum runs over the reciprocal lattice of the rectangular lattice defined by the two-site unit cell of figure \ref{fig:Unit cells}(a).
$M(\bf{q})$ in Eq.~(\ref{eq:Energy to 8th order}) is the $2 \times 2$ matrix
\begin{equation}
M(\bf{q})=\left(
\begin{array}{cc}
 -2\cos q_z &1+e^{-i2q_x} \\
 1+e^{i2q_x} & 2\cos q_z
\end{array}\right).
\end{equation}
The eigenvalues of $M(\bf{q})$ are
\begin{equation}\label{eq:Eigenvalues of M(q)}
 \lambda_{\pm}(q_x,q_z)=\pm\sqrt{2}\sqrt{2+\cos{2 q_x}+\cos{2q_z}}.
\end{equation}
The largest eigenvalues are given by
$\lambda_+(0,0)=\lambda_+(0,\pm\pi)=2\sqrt{2}$. Therefore, the quadratic form
obtained by truncating Eq.~(\ref{eq:Energy to 8th order}) to the second
order in $n^z$ is positive definite
as long as $\Gamma/J>2\sqrt{2}$. This defines the critical value of the
transverse field $\Gamma_c=2\sqrt{2}J$ at which the fully polarized state
becomes unstable. The wavevectors of the critical modes, ${\bf q}_A=(0,0)$
and ${\bf q}_B=(0,\pm\pi)$, indicate that an instability occurs towards a phase with a unit cell which is twice as large and contains four sites, see
Fig.~\ref{fig:Unit cells}(b).

These critical modes are real and can be parametrized as
\begin{equation}
\begin{array}{c}
 (n_{{\bf q}_A,1}^z,n_{{\bf q}_A,2}^z)=n_A^z {\bf u}_A^{(+)} ,
\\[3mm]
 (n_{{\bf q}_B,1}^z,n_{{\bf q}_B,2}^z)=n_B^z {\bf u}_B^{(+)} ,
\end{array}
\end{equation}
where the real coefficients $n_A^z$ and $n_B^z$ are their amplitudes and
\begin{equation}
\begin{array}{lll}
 {\bf u}_A^{(+)}=\left(\sin{\frac{\pi}{8}},\cos{\frac{\pi}{8}}\right) , &\quad &
 {\bf u}_B^{(+)}=\left(\cos{\frac{\pi}{8}},\sin{\frac{\pi}{8}}\right)
\end{array}
\end{equation}
are the normalized eigenvectors of $M({\bf q}_A)$ and $M({\bf q}_B)$
associated to the eigenvalue $+2\sqrt{2}$. Since the degeneracy of the
two critical modes is not related to symmetry, one can expect it to be
removed in the higher orders of the expansion of the classical energy
(\ref{eq:E classical}).

The minimization of
\begin{equation}
 \mathcal{E}_0^{(4)}=
\frac{\Gamma-\Gamma_c}{4}\left[(n_A^z)^2+(n_B^z)^2\right]+
\frac{3\Gamma}{64}\left[(n_A^z)^2+(n_B^z)^2\right]^2\;,
\end{equation}
the sum of the second- and fourth-order contributions of the critical
modes to Eq.~(\ref{eq:Energy to 8th order}), fixes only the combination
\begin{equation}                                 \label{eq:n^z squared}
(n_A^z)^2+(n_B^z)^2=\frac{8}{3\Gamma}(\Gamma_c-\Gamma)\;,
\end{equation}
whereas the ratio between $n_A^z$ and $n_B^z$ remains undefined. It
follows from Eq. (\ref{eq:n^z squared}) that, in the leading order,
\makebox{$n_A^z\sim n_B^z \sim (\Gamma_c -\Gamma)^\frac{1}{2}$} and
$\mathcal{E}_0^{(4)}\sim (\Gamma_c-\Gamma)^2$.

Taking into account the sixth-order contributions to the energy one
obtains
\begin{equation}
 \mathcal{E}_0^{(6)}= \mathcal{E}_0^{(4)}+
\frac{5\Gamma}{256}\left[(n_A^z)^2+(n_B^z)^2\right]^3.
\end{equation}
As $\mathcal{E}_0^{(4)}$, $\mathcal{E}_0^{(6)}$ depends only on
$(n_A^z)^2+(n_B^z)^2$. The dependence on the relative strength of the two
amplitudes appears only in the eighth-order contribution to the energy
\begin{equation}\label{eq:E08}
\begin{array}{lll}
 \mathcal{E}_0^{(8)}&=& \mathcal{E}_0^{(6)}+
\frac{5\Gamma}{8192}
\left[17(n_A^z)^8+84(n_A^z)^6(n_B^z)^2\right.\\
& &\left.+70(n_A^z)^4(n_B^z)^4+
84(n_A^z)^2(n_B^z)^6+17(n_B^z)^8\right].
\end{array}
\end{equation}

The minimization of Eq.~(\ref{eq:E08}) fixes the ratio $n_A^z/n_B^z$. This
is readily seen when parametrizing the amplitudes of the critical modes as
\begin{equation}
 \begin{array}{ccc}
  n_A^z=n_z\cos{\phi},& \quad & n_B^z =n_z\sin{\phi}.
 \end{array}
\end{equation}
With this choice, the sum of the second- to eighth-order contributions to
the energy becomes
\begin{equation}\label{eq:E08 new parametrization}
 \mathcal{E}_0^{(8)}=\frac{\Gamma-\Gamma_c}{4}n_z^2
                 +\frac{3\Gamma}{64}n_z^4
             +\frac{5\Gamma}{256}n_z^6
         +\frac{5\Gamma}{8192}n_z^8\frac{(35-\cos{8\phi})}{2}
\end{equation}
which, for any value of $n_z$, is minimal when \makebox{$\cos{8\phi}=1$}. This
selects eight degenerate solutions $\phi=p\pi/4$ with \makebox{$p=0,\,1,\ldots,7$}. To the lowest order in $\Gamma_c-\Gamma$, the corresponding real space spin configurations are
\begin{equation}\label{eq:Eight columnar solutions}
 \begin{array}{l}
\displaystyle n_{1}^z=n_z\sin\left[\frac{\pi}{8}\left(2p+1\right)\right]\,, \quad
\displaystyle n_{2}^z=n_z\sin\left[\frac{\pi}{8}\left(2p+3\right)\right]\,,\\[3mm]
\displaystyle n_{3}^z=n_z\sin\left[\frac{\pi}{8}\left(2p+5\right)\right]\,, \quad
\displaystyle n_{4}^z=n_z\sin\left[\frac{\pi}{8}\left(2p+7\right)\right]\,,
 \end{array}
\end{equation}
where $n_z=\sqrt{8(\Gamma_c-\Gamma)/3\Gamma_c}$, and where the indices that run from $1$ to $4$ keep track of the four sublattices
[see Fig.~\ref{fig:Unit cells} (b)].
In terms of the dimer density distribution, the eight solutions correspond to the four
states with so-called columnar structure. The example of such a
state presented in Fig.~\ref{fig:Dimer Patterns}(b) corresponds to the solution with $p=0$ in Eq.~(\ref{eq:Eight columnar solutions}). These four
columnar coverings are equivalent, that is, they are related to each other by
symmetry. Each of them corresponds to two different values of $\phi$ which
differ by $\pi$. In terms of classical spins the shift of $\phi$ by $\pi$
corresponds to reflecting all spins with respect to the $xy$ plane.

This approach is based on the assumption that all other modes contribute
to the energy expansion only in the higher orders. However, it is known
from the analysis of the \makebox{FFTFIM} on the honeycomb lattice, \cite{Coletta11}
that when it is necessary to go beyond the fourth order of the
Ginzburg-Landau expansion, by restricting the calculation just to the
subspace of the critical modes of $M({\bf q})$, one may miss some contributions
which are equally essential for the structure selection. We shall now show that in the considered model,
some second-, fourth- and sixth-order terms involving noncritical (gapped)
modes also make contributions that are essential for determining the
optimal value of $\phi$.

The momenta of the critical modes enforce a four-sublattice  structure of
the ground state which allows for the excitation of other modes compatible
with the same periodicity. Since the only wave vectors consistent with
this periodicity are those of the critical modes, the momenta of the relevant
noncritical modes will also be ${\bf q}={\bf q}_A$ and ${\bf q}={\bf
q}_B$. The most important terms coupling the critical modes with extra
modes are expected to be linear in the amplitudes of these extra modes and
of a higher order in the amplitudes of critical modes.

Let us denote the Fourier coefficients associated to these extra modes by
\begin{equation}
\begin{array}{c}
 (\bar{n}_{{\bf q}_A,1}^z,\bar{n}_{{\bf q}_A,2}^z)=\bar{n}_A^z
{\bf u}_A^{(-)},
\\[3mm]
 (\bar{n}_{{\bf q}_B,1}^z,\bar{n}_{{\bf q}_B,2}^z)=\bar{n}_B^z
{\bf u}_B^{(-)},
\end{array}
\end{equation}
where the real coefficients $\bar{n}_A^z$ and $\bar{n}_B^z$ are their
amplitudes and
\begin{equation}
 \begin{array}{lll}
  {\bf u}_A^{(-)}=\left(-\cos{\frac{\pi}{8}},\sin{\frac{\pi}{8}}\right)&,\quad&
  {\bf u}_B^{(-)}=\left(-\sin{\frac{\pi}{8}},\cos{\frac{\pi}{8}}\right)
 \end{array}
\end{equation}
are the normalized eigenvectors of $M({\bf q_A})$ and $M({\bf q_B})$
associated to the eigenvalue $-2\sqrt{2}$.
The terms in the energy functional to sixth-order that are linear and harmonic
in the extra modes are
\begin{widetext}
\begin{equation}\label{eq:sub critical modes}
\begin{array}{ll}
 \mathcal{E}_1^{(6)}&=\displaystyle\frac{\Gamma_c+\Gamma}{4}\left[
(\bar{n}_A^z)^2+(\bar{n}_B^z)^2\right]
 +\frac{\Gamma}{16}\left[(n_A^z)^3\bar{n}_A^z
+3(n_A^z)^2 n_B^z \bar{n}_B^z
-3(n_B^z)^2 n_A^z \bar{n}_A^z
-(n_B^z)^3\bar{n}_B^z
\right]
\\[3mm]
& \displaystyle+\frac{3\Gamma}{32}\left[
(n_A^z)^2(\bar{n}_A^z)^2
+3(n_A^z)^2(\bar{n}_B^z)^2
+3(n_B^z)^2 (\bar{n}_A^z)^2
+(n_B^z)^2(\bar{n}_B^z)^2
+4 n_A^zn_B^z \bar{n}_A^z \bar{n}_B^z
\right] \\[3mm]
& \displaystyle+\frac{3\Gamma}{64}\left[(n_A^z)^5 \bar{n}_A^z
+5(n_A^z)^4 n_B^z \bar{n}_B^z -5(n_B^z)^4 n_A^z \bar{n}_A^z
-(n_B^z)^5\bar{n}_B^z
\right].
\end{array}
\end{equation}
\end{widetext}

The dominant contribution to the amplitudes of the subcritical modes is
captured by the variation of the first line of Eq.~(\ref{eq:sub critical
modes}) with respect to $\bar{n}_A^z$ and $\bar{n}_B^z$. This yields
\begin{equation}\label{eq:Bar n}
 \begin{array}{lll}
  \bar{n}_A^z&\approx&\frac{1}{16} \left[-(n_A^z)^3+3(n_B^z)^2n_A^z\right],\\[3mm]
  \bar{n}_B^z&\approx&\frac{1}{16} \left[(n_B^z)^3-3(n_A^z)^2n_B^z\right],\\
 \end{array}
\end{equation}
indicating that the lowest order contribution to these quantities is
$\bar{n}_A^z\sim \bar{n}_B^z\sim\left(\Gamma_c-\Gamma\right)^\frac{3}{2}$. Note that in
Eqs.~(\ref{eq:Bar n}) $\Gamma$ has been replaced by $\Gamma_c$. This has
been done because we are focused only on the leading order contributions
in $\Gamma_c-\Gamma$. Injecting solution (\ref{eq:Bar n}) into
(\ref{eq:sub critical modes}) one obtains
\begin{equation}
 \mathcal{E}_1^{(6)}=-\frac{\Gamma_c n_z^6}{512}
+\frac{3\Gamma_c n_z^8(-8+\cos{8\phi})}{8192}.
\end{equation}

Hence $\mathcal{E}_1^{(6)}$ is of same order as $\mathcal{E}_0^{(8)}$.
Combining the terms which are of order $(\Gamma-\Gamma_c)^4$ coming from
$\mathcal{E}_0^{(8)}$ and $\mathcal{E}_1^{(6)}$ and which depend on $\phi$
yields
\begin{equation}\label{eq:Alpha dependant terms}
\frac{1}{8192}\frac{\Gamma_c n_z^8}{2}\cos{8\phi}\;.
\end{equation}
Since the prefactor of $\cos{8\phi}$ in this expression is positive, the
minimum of the energy is achieved when $\cos{8\phi}=-1$. This defines
eight degenerate solutions $\phi=(2p'+1)\pi/8$ with $p'=0,\,1,\ldots,7$.
The corresponding real space spin configurations can be obtained replacing
$p$ by $p'+1/2$ in Eq.~(\ref{eq:Eight columnar solutions}).
However, the energetic stability of the solution $p=p'+1/2$
in comparison with other solutions is ensured by the form of the
higher-order corrections to (\ref{eq:Eight columnar solutions}) which are proportional to $n_z^3$.
All eight degenerate solutions corresponding to $p=p'+1/2$ have one of the spins in the four-sublattice structure aligned with the transverse field. All
nearest neighbors to the polarized spin have the same $n^x$ component
which is greater than that of next nearest neighbors
[Fig~\ref{fig:Dimer Patterns}(a) presents an example of a state with such
a structure corresponding to the solution with $p'=0$].
In the dimer representation, these solutions correspond to
four equivalent plaquette structures which can be obtained from one
another by translation. The rotational symmetry of the plaquette state is
$C_4$ whereas that of the columnar state is $C_2$.

In this approach only the leading contributions to the amplitudes
$n_A^z,n_B^z,\bar{n}_A^z$ and $\bar{n}_B^z$ in powers of $\Gamma_c-\Gamma$
are considered. Hence, one may wonder if the higher order corrections to
these amplitudes may lead to other $\phi$ dependent terms relevant for the
structure selection. We verified that this is not the case and that to
order $(\Gamma_c-\Gamma)^4$ all $\phi$ dependent terms are captured by the
expression in Eq.~(\ref{eq:Alpha dependant terms}). This was done by
observing that the energy difference between the columnar and plaquette
structures obtained by numerical calculation of their energies in the
vicinity of $\Gamma_c$ is equal to the prefactor in expression
Eq.~(\ref{eq:Alpha dependant terms}) multiplied by two since for the columnar state
$\cos8\phi=1$ and for the plaquette state $\cos8\phi=-1$.

From this calculation we deduce that just below the critical field
$\Gamma_c$ the classical solution with the lowest energy has the plaquette
structure and that the energy difference between the plaquette and
columnar states is of the order of $(\Gamma_c-\Gamma)^4/J^3$ per site. In
the next section we analyze the properties of the ground states in the
opposite limit of weak fields.


\subsection{Weak field case} \label{sec:Weak field}

In the weak field limit, the structure of the ground states can be also
studied analytically. This allows one to understand in simple terms why
part of the spins are completely polarized along the field. At zero field, the
energy of a single square plaquette is given by
\begin{equation}\label{SinglePl}
    E = \sum_{m=1}^{p}J_{m,m+1}n^z_m n^z_{m+1}
\end{equation}
where subscript $m=1,\ldots,p$ numbers the spins along the perimeter of
the plaquette, $p=4$  and $n^z_{p+1}\equiv n^z_1$.

It has been proven in Ref.~\onlinecite{Coletta11} that for $p=6$ the
minimum of this expression is achieved in the states in which one of the
spins (for example, the one with $m=p$) has an arbitrary orientation,
whereas the remaining $p-1$ spins are all directed along the $z$ axis with
\begin{equation}\label{theRule}
n^z_{m+1}=-{\rm sign}(J_{m,m+1})n^z_m=\pm 1\,,
\end{equation}
which corresponds to minimizing the energy on each of the bonds connecting
them (as in an open chain of length \makebox{$p-1$}). This proof does not
rely on the particular value of $p$ and is applicable for any $p\geq 3$,
in particular for $p=4$. The ground state energy of a single plaquette is
therefore equal to \makebox{$-J(p-2)$}, which for $p=4$ gives $-2J$.

On the honeycomb lattice ($p=6$), it is possible to minimize
simultaneously the energy of each of the three plaquettes sharing a given
site only if the spin on this site is directed along $z$ ($n^z_j=\pm 1$).
For this reason, the ground state manifold of the model
(\ref{eq:Generalized Classical Hamiltonian}) with $\Gamma = 0$ coincides
with that of the discrete Ising model with the same Hamiltonian but
$n^z_j=\pm 1$.~\cite{Coletta11}

In contrast to that, on the square lattice, the ground state manifold of
the model (\ref{eq:Generalized Classical Hamiltonian}) with  $\Gamma = 0$
is essentially wider than that of its discrete version. Different grounds
states can be obtained by taking a ground state of the discrete model and
rotating  in an arbitrary way some of the spins for which the sum
$\sum_{j}J_{ij}n^z_{j}$ describing their interaction with the neighboring
spins is equal to zero. However, two rotated spins cannot be the nearest
neighbors of each other.

The application of a weak transverse field strongly suppresses the
degeneracy of the ground states. It is evident that in order to decrease
the energy, all spins which in the absence of the field are free to rotate
now have to be directed along the field. This gives a negative
contribution to the energy equal to $\Gamma$ times the number of such spins.
Therefore, in the states minimizing the energy this number has to be as
large as possible. Since each polarized spin sits in the middle of a
$2\times 2$ cell along whose perimeter the spins have to be parallel (or
antiparallel) to the $z$ axis, the maximal fraction of spins polarized by
the field is equal to one quarter. In such a case, on each plaquette one
of the four spins is polarized by the field. In particular, this can be
realized in the state with the four-sublattice structure.
In the next section we show that for any field the global minimum of
energy can also be achieved in the framework of the four-sublattice
ansatz.

\subsection{Reduction to the single-plaquette problem}\label{sec:four-sublattice structure}

The aim of this section is to show that the minimum of the energy for
any field $\Gamma$ can be achieved in a state with the four-sublattice
structure. To demonstrate this, it is convenient to rewrite the classical
energy (\ref{eq:Generalized Classical Hamiltonian}) as
\begin{equation}              \label{eq:Energy as sum of plaquettes}
E=\sum_{\alpha}E_\alpha,
\end{equation}
where
\begin{equation}                            \label{eq:Ealpha}
\begin{array}{ll}
E_\alpha=&\displaystyle-\frac{J}{2}\left[n_{j_1(\alpha)}^z
n_{j_2(\alpha)}^z +n_{j_2(\alpha)}^z n_{j_3(\alpha)}^z+n_{j_3(\alpha)}^z
n_{j_4(\alpha)}^z
-n_{j_4(\alpha)}^z n_{j_1(\alpha)}^z\right]\\
&\displaystyle-\frac{\Gamma}{4}\sum_{m=1}^4\sqrt{1-[n_{j_m(\alpha)}^z]^2},
\end{array}
\end{equation}
the index $\alpha$ runs over all four-site plaquettes of the lattice, and
$j_m(\alpha)$ denotes the site belonging to plaquette $\alpha$ and to
sublattice number $m$ [$m=1,2,3,4$, see Fig. \ref{fig:Unit cells}(b)].
Each term $E_\alpha$ depends only on four variables $n_{j_m(\alpha)}^z$
which are associated to the four sites belonging to plaquette $\alpha$.
Below $E_\alpha$ is often called the energy of plaquette $\alpha$.

It is evident that if it is possible to minimize simultaneously all terms
in the sum (\ref{eq:Energy as sum of plaquettes}), this will give the
absolute minimum of energy. Since Eq. (\ref{eq:Ealpha}) has exactly the
same structure for all plaquettes, this aim is easily achieved by
minimizing $E_\alpha$ for a single plaquette and then assuming that the
state has the four-sublattice structure in which all variables $n_j^z$
defined on the sites belonging to sublattice $m$ have the same
value.

The reasoning above does not prove that {\em all} ground states have to
have the four-sublattice structure. To check if this is really so it is
necessary first to find what spin configurations minimize $E_\alpha$,
which is the topic of the next section.

\subsection{Single plaquette energy minimization}\label{sec:Single Plaquette}

In Sec. \ref{sec:Weak field} we have shown that in the limit $\Gamma\to 0$ the
energy is minimized when on each plaquette one of the spins is fully
polarized along the field. On the other hand, the results of Sec.
\ref{sec:Ginzburg-Landau} suggest that the same property holds also when
$\Gamma$ approaches $\Gamma_c$. It seems plausible that the state with
such a structure minimizes the energy for any $\Gamma\in(0,\Gamma_c)$. To
check this, let us first find the explicit form of the spin configuration
minimizing the energy of a single plaquette $E_\alpha$ defined by Eq.
(\ref{eq:Ealpha}) under the assumption that one of the four spins is fully
polarized. Let us denote by $i$ the site where the spin is polarized along
the field  [that is, $n_i=(1,\;0,\;0)$], by $j$ and $k$ its two neighbors
and by $l$ the site diagonally opposite to $i$. Under this assumption the
energy of a single plaquette can be written as
\begin{equation}\label{eq:Energy single plaquette one spin polarized}
 E_\alpha 
 =\frac{1}{2}n_l^z\left(J_{lj}n_j^z+J_{lk}n_k^z\right)
 -\frac{\Gamma}{4}\left(1+n_j^x+n_k^x+n_l^x\right).
\end{equation}
The spin configuration minimizing (\ref{eq:Energy single plaquette one spin polarized}) at any field is given by:
\begin{equation}\label{eq:Sol one Plaquette}
 (n_l^z)^2=\frac{1-\frac{1}{4}\left(\frac{\Gamma}{2J}\right)^4}{1+\left(\frac{\Gamma}{4J}\right)^2}, \quad
 (n_j^z)^2=(n_k^z)^2=\frac{1-\frac{1}{4}\left(\frac{\Gamma}{2J}\right)^4}{1+\left(\frac{\Gamma}{2J}\right)^2}< (n_l^z)^2,
\end{equation}
with the signs of $n_j^z$ and $n^z_k$ determined by the sign of $n^z_l$,
\begin{equation}\label{eq:Sol one Plaquette signs}
 \textrm{sign}(n_j^z)=-\textrm{sign}(J_{jl}n_l^z)\;,~~~~~~~
 \textrm{sign}(n_k^z)=-\textrm{sign}(J_{kl}n_l^z)\;.
\end{equation}
The solution described by Eqs. (\ref{eq:Sol one Plaquette}) and
(\ref{eq:Sol one Plaquette signs}) is valid regardless of the position of
the fully polarized spin with respect to the antiferromagnetic bond as
long as $j$ and $k$ denote the sites neighboring the polarized spin and
$l$ the site diagonally opposite to it.

It follows from Eqs.~(\ref{eq:Sol one Plaquette}) and (\ref{eq:Sol one
Plaquette signs}) that in such a state the force acting on
the spin at site $i$ is equal to zero,
\begin{equation}\label{}
   J_{ij}n_j^z+J_{ik}n_k^z =0\,.
\end{equation}
This implies that the same energy can be obtained whatever the value of
 $n_i^z$, and not only for $n_i^z=0$. We have verified with the help of a numerical
minimization of $E_\alpha$ as a function of four variables that the
solutions found above are not local but global minima of
$E_\alpha$. In total, for a given plaquette there are eight spin
configurations minimizing its energy, which are related to each other by
symmetries. The factor four comes from the possibility to choose a site at
which the spin is fully polarized and the additional factor two comes from
the possibility to choose the sign of $n^z_l$.

\subsection{Classical ground states 
and their dimer representation}\label{subsec:Classical ground states and their dimer representation}

After finding eight spin configurations minimizing the energy of a single
plaquette we can immediately construct eight ground states of the model on
the infinite square lattice. As it has been already mentioned  in Sec.
\ref{sec:four-sublattice structure}, this can be achieved by choosing one
of these configuration for one particular plaquette and after that
assuming that the ground state has a four-sublattice structure, that is,
on each of the four sublattices [see Fig. \ref{fig:Unit cells}(b)] all spins
have the same orientation. It turns out to be impossible to construct any other
state with the same energy, because on each of the bonds the two spins
belonging to it always have different orientations. Therefore, the choice
of one of the eight configurations on one plaquette uniquely determines
which configurations have to be chosen on neighboring plaquettes, and so
on, which reproduces nothing else but a state with the four-sublattice
structure.  Accordingly, the ground state manifold is restricted to the
eight four-sublattice states.

In terms of the gauge-invariant variables $d_{ij}$ defined by Eq.
(\ref{eq:Dimer Density}) (which in the quantum case can be interpreted as
dimer densities on the bonds of the dual lattice crossing the
corresponding bonds of the original lattice) these ground states
correspond to the plaquette dimer structure
exhibiting a $C_4$ rotational symmetry. In this structure the dimer
densities on the bonds of the dual lattice surrounding the sites of the
original lattice on which the spins are fully polarized by the field are
all equal to $1/2$, while the dimer densities of all remaining bonds of
the dual lattice are again all equal to each other  but are smaller than
$1/2$. Hence the bonds with the highest dimer densities form a regular
pattern of square plaquettes, as shown in Fig. \ref{fig:Dimer
Patterns}(a), which explains the origin of the widely used term
``plaquette phase". The plaquette pattern is characterized by a four-fold
degeneracy related to translations, however each pattern corresponds to
two different ground states in terms of classical spins. These two states
are transformed into each other by a reflection of all the spins with respect
to $xy$ plane (in other terms, by changing the signs of all $n_j^z$).

\begin{figure}
 \centering
\includegraphics[width=\columnwidth]{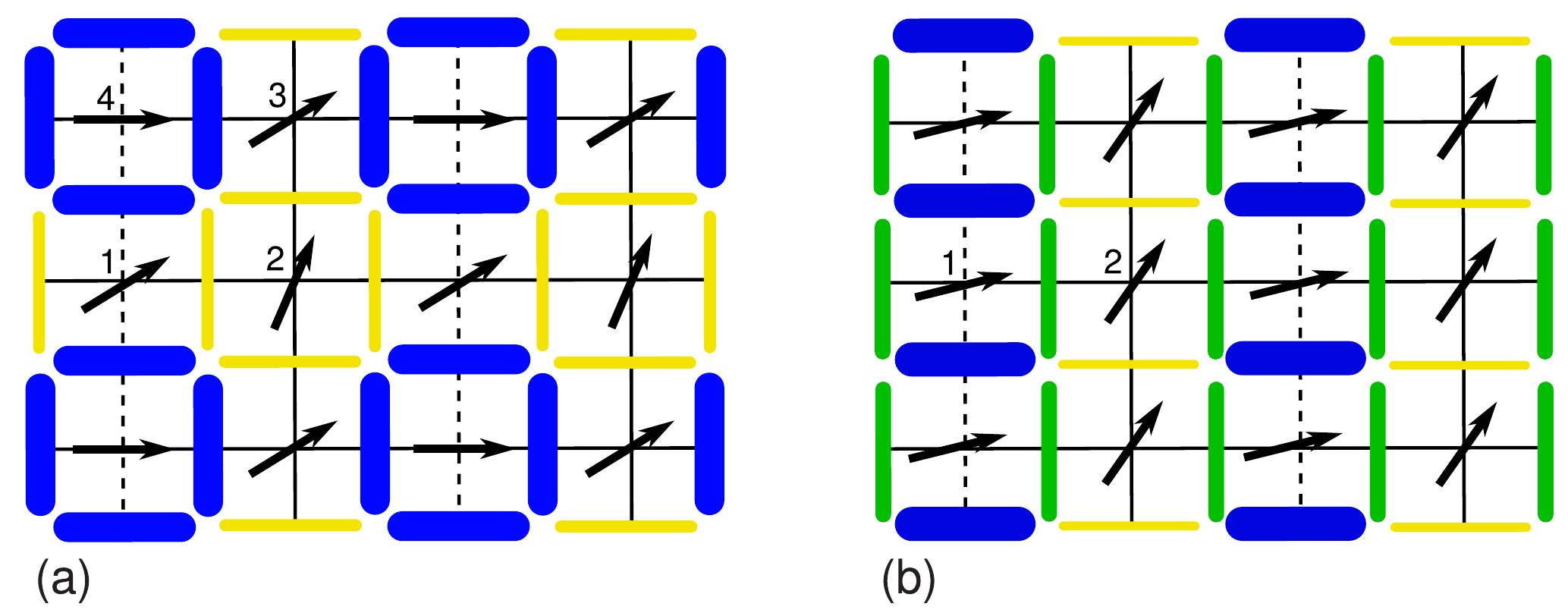}
\caption{(Color Online) Examples of plaquette (a) and columnar (b) dimer patterns.
The thickness of the bonds is proportional to the dimer density $d_{ij}$.
The corresponding classical spin configuration of the FFTFIM is also depicted.}
\label{fig:Dimer Patterns}
\end{figure}

The columnar state corresponds to the dimer density pattern of the type
depicted in Fig.~\ref{fig:Dimer Patterns} (b). In the framework of the
classical model, this state appears if the analytical analysis in the
vicinity of $\Gamma_c$ takes into account only the critical modes. The
columnar pattern has a four-fold degeneracy with one factor two related to
translations and another one to rotations by 90 degrees. Naturally, in
terms of spins, the degeneracy is doubled, because gauge-invariant
variables $d_{ij}$ are not sensitive to a simultaneous change of signs of
all $n^z_j$.  All eight spin realizations of the columnar state have a
periodicity compatible with the four-sublattice structure [see Fig.
\ref{fig:Unit cells}\,(b)]. However, for each of them, it is possible to choose
the gauge in such a way that the spin configuration becomes compatible with a
larger translation group [that is, has a two-sublattice structure].
In particular, the choice of gauge depicted in Fig.~(\ref{fig:Unit cells}) is the one
realizing the doubling of the translation group of the columnar solution
represented in Fig.~\ref{fig:Dimer Patterns} (b).
For other columnar structures this gauge is obtained from that of Fig.~(\ref{fig:Unit cells})
by a translation of one lattice parameter in the $x$ direction, or by a rotation by $\pi/2$.

Within the 4-sublattice ansatz the columnar solution corresponds to a
saddle point of the classical energy lying between two plaquette states.
More precisely, in the phase space of four-sublattice structures there
exists an almost degenerate circle of low energy states which has eight
equivalent minima (the plaquette states) and between them eight degenerate
maxima corresponding to the columnar structures.
In the vicinity of $\Gamma_c$, this family of low energy states can be
parametrized by Eqs.~(\ref{eq:Eight columnar solutions}) treating $p$
as a continuous variable.
Insofar the emergence of the eight-fold degeneracy in the \makebox{FFTFIM} was discussed \cite{MoessnerSondhi01}
only in relation with its appearance in the framework of the
Ginzburg-Landau expansion for the free energy of the classical
three-dimensional version of this model.\cite{BlankschteinMaBerker88} Our
analysis has revealed that such a degeneracy has an even more evident
origin.

The energy barrier separating ``neighboring" plaquette states (that is,
the difference in energy between the plaquette and columnar states) is
always relatively small. In the vicinity of $\Gamma_c$ its value (per
site) is of the order of $(\Gamma_c-\Gamma)^4/J^3$ and is much smaller
than the characteristic energies of the plaquette and columnar state
(counted off from that of the paramagnetic one) which both are of the
order of $(\Gamma_c-\Gamma)$. In the low field limit the height of the
barrier is of the order of $\Gamma$ and is small in comparison with $J$,
the energy scale characterizing the manifold of the four-sublattices
states.
Between the analytically tractable limits, we verified numerically that
the energy difference between the columnar and plaquette states never exceeds
$0.013J$ per site. This value is achieved at $\Gamma\approx0.3J$, where the
characteristic energy scale is still of the order of $J$. In the following
section we show that fluctuations around the considered periodic
structures change the classical picture and stabilize the columnar
solution over the plaquette one.

\section{Semiclassical approach}
\label{sec:Semiclassical approach}
\subsection{Linear spin-wave approximation}
Having discussed the classical phase diagram of the model we now examine the effect of quantum fluctuations on the competition between the plaquette and columnar states.
Quantum fluctuations are investigated in the context of the large
$S$ expansion (spin-wave approximation).
Without loss of generality we consider, out of the eightfold degenerate plaquette solutions, the one having the same orientation of the spins on the sublattices $1$ and $3$ and fully polarized spins on sublattice $4$ [see Fig.~\ref{fig:Dimer Patterns}(a)], with the signs of $n_1^z=n_3^z$ and $n_2^z$ chosen to be positive. The columnar solution used in the calculations is the one having the simplest structure in the gauge used in this work and
consisting of just two sublattices, see Fig.~\ref{fig:Dimer Patterns}(b). These structures are the starting classical solutions used in our spin-wave calculation.

The first step in the construction of the spin-wave expansion consists in rotating the spin operators on each site around the $y$ axis,
\begin{equation}\label{eq:LocalSpin Rotation}
\begin{array}{lll}
 S^x_{j} & = & n^z_{m(j)} S_{j}^{x^\prime}+n^{x}_{m(j)} S^{z^\prime}_{j}\,, \\[2mm]
 S^y_{j} & = & S_{j}^{y^\prime}\,, \\[2mm]
 S^z_{j} & = & -n^x_{m(j)} S_{j}^{x^\prime}+n^z_{m(j)} S_{j}^{z^\prime}\,,
\end{array}
\end{equation}
Here $m(j)$ denotes the number of the sublattice to which site $j$ belongs, whereas $n_m^x$ and $n_m^z$ are the two components of the classical spin lying in the $xz$ plane and belonging to the $m^\textrm{th}$ sublattice.
The purpose of this rotation is to achieve a situation where on each site the axis $z^\prime$ is aligned with the classical spin. After that the spin operators in the rotated frame are mapped to bosons via the standard
Holstein-Primakoff (HP) transformation \cite{Holstein-Primakoff} according to
\begin{equation}\label{eq:HP transformation}
 \begin{array}{c}
S_{j}^{x^\prime}+iS_{j}^{y^\prime}=\displaystyle \sqrt{2S}\sqrt{1-\frac{a_{j}^\dagger a_{j}^\dagga}{2S}}a_{j}^\dagga\,, \\
S_{j}^{x^\prime}-iS_{j}^{y^\prime}=\displaystyle \sqrt{2S}a_{j}^\dagger\sqrt{1-\frac{a_{j}^\dagger a_{j}^\dagga}{2S}}\,, \\[3mm]
S_{j}^{z^\prime}=S-a_{j}^\dagger a_{j}^\dagga\,.
 \end{array}
\end{equation}
The spin-wave approximation is based on replacing the square root in Eqs.~(\ref{eq:HP transformation}) by its expansion in powers of $a_{j}^\dagger a_{j}^\dagga/S$,
the number of HP particles divided by the value of the spin being the small expansion parameter.
When truncating the expansion to harmonic level, the original Hamiltonian (\ref{eq:Generalized Quantum Hamiltonian}) expressed in terms of Holstein-Primakoff bosons
can be split intro three contributions
\begin{equation}
\mathcal{H}=\mathcal{H}^{(0)}+\mathcal{H}^{(1)}+\mathcal{H}^{(2)} ,
\end{equation}
where $\mathcal{H}^{(0)}$ is the classical energy of the system, $\mathcal{H}^{(1)} \propto S^{-\frac{1}{2}}$
and $\mathcal{H}^{(2)}\propto S^{-1}$ respectively contain only terms which are linear and quadratic in bosonic operators.
Furthermore, the coefficients in $\mathcal{H}^{(1)}$ are proportional to the derivatives of the classical energy with respect to the spin orientations.
For this reason, $\mathcal{H}^{(1)}=0$ both for the plaquette solution, which is the ground state of the classical energy, and for the columnar solution, which is a saddle point of the classical energy.

After Fourier transformation the quadratic bosonic Hamiltonian of the plaquette structure can be expressed as
\begin{equation}\label{eq:Fluctuation H plaquette}
 \mathcal{H}^P=NE^P+\sum_{\bf q}\left[\vec{a}_{\bf q}^\dagger \bar{H}_{\bf q}^P\vec{a}_{\bf q}+\Delta\right] ,
\end{equation}
where $E^P$ is the classical energy per site of this configuration. Its transverse field dependence
is given by
\begin{equation}
 E^P=-\frac{\Gamma}{4}-\frac{\sqrt{(4J^2+\Gamma^2)(16J^2+\Gamma^2)}}{8J}\,.
\end{equation}
In Eq.~(\ref{eq:Fluctuation H plaquette}) $\vec{a}_{\bf q}^\dagger=\left(a_{{\bf q},1}^\dagger\ldots,a_{{\bf q},4}^\dagger,a_{-{\bf q},1}^\dagga\ldots,a_{-{\bf q},4}^\dagga\right)$
and $\bar{H}_{\bf q}^P$ is the $8\times8$ matrix
\begin{equation}
 \bar{H}_{\bf q}^P=
 \left(\begin{array}{cccccccc}
  A & E_{\bf q}^\star & 0 & D_{\bf q}^\star & 0 & E_{\bf q}^\star & 0 & D_{\bf q}^\star \\
  E_{\bf q} & C & G_{\bf q}^\star & 0 & E_{\bf q} & 0 & G_{\bf q}^\star & 0 \\
  0 & G_{\bf q} & A & F_{\bf q} & 0 & G_{\bf q} & 0 & F_{\bf q} \\
  D_{\bf q} & 0 & F_{\bf q}^\star & B & D_{\bf q} & 0 & F_{\bf q}^\star & 0 \\
  0 & E_{\bf q}^\star & 0 & D_{\bf q}^\star & A & E_{\bf q}^\star & 0 & D_{\bf q}^\star \\
  E_{\bf q} & 0 & G_{\bf q}^\star & 0 & E_{\bf q} & C & G_{\bf q}^\star & 0\\
  0 & G_{\bf q} & 0 & F_{\bf q} & 0 & G_{\bf q} & A & F_{\bf q} \\
  D_{\bf q} & 0 & F_{\bf q}^\star & 0 & D_{\bf q} & 0 & F_{\bf q}^\star & B \\
 \end{array}
\right),
\end{equation}
with coefficients
\begin{equation}
 \begin{array}{ll}
  A=\displaystyle\frac{1}{2S}\left(2J n_1^z n_2^z+\Gamma n_1^x\right) , &
  B=\displaystyle\frac{\Gamma}{2S},  \\[3mm]
  C=\displaystyle\frac{1}{2S}\left(4J n_1^z n_2^z+\Gamma n_2^x\right), &
  D_{\bf q}=\displaystyle\frac{J}{4S}n_1^x\left(1+e^{i2q_z}\right), \\[3mm]
  E_{\bf q}=\displaystyle-\frac{J}{4S}n_1^xn_2^x\left(1+e^{i2q_x}\right), &
  F_{\bf q}=\displaystyle-\frac{J}{4S}n_1^x\left(1+e^{i2q_x}\right), \\[3mm]
  G_{\bf q}=\displaystyle-\frac{J}{4S}n_1^xn_2^x\left(1+e^{i2q_z}\right). &
 \end{array}
\end{equation}
In the above expressions, $n_m^x$ and $n_m^z$ with $m=1,2$ are the components of the classical spins in the plaquette state considered
in this spin-wave calculation, see Fig.~\ref{fig:Dimer Patterns}(a).
Their values are given by Eqs.~(\ref{eq:Sol one Plaquette})
with the identification $j=k=1$ and $l=2$.
The term $\Delta$ in Eq.~(\ref{eq:Fluctuation H plaquette}) is defined by $\Delta=-\left(2A+B+C\right)$.

Similarly, the quadratic bosonic Hamiltonian of the columnar structure is of the form
\begin{equation}\label{eq:Fluctuation H columnar}
 \mathcal{H}^C=NE^C+\sum_{\bf q}\left[\vec{a}_{\bf q}^\dagger \bar{H}_{\bf q}^C\vec{a}_{\bf q}+\tilde{\Delta}_{\bf q}\right] ,
\end{equation}
where $E^C$ is the classical energy per site of the columnar structure.
The energy per site of the structure having the symmetries of the columnar state [Fig.~\ref{fig:Dimer Patterns} (b)] 
takes the form
\begin{equation}\label{eq:E columnar}
E^C({\bf n}_1,{\bf n}_2)=
-Jn_1^zn_2^z-\frac{J}{2}\left(n_2^z\right)^2+\frac{J}{2}\left(n_1^z\right)^2
-\frac{\Gamma}{2}(n_1^x+n_2^x)\,,
\end{equation}
and, accordingly,
$E^C$ in Eq.~(\ref{eq:Fluctuation H columnar}) denotes the minimum of
$E^C({\bf n}_1,{\bf n}_2)$ with respect to ${\bf n}_1$ and ${\bf n}_2$.
In Eq.~(\ref{eq:Fluctuation H columnar}) \makebox{$\vec{a}_{\bf q}^\dagger=\left(a_{{\bf q},1}^\dagger,a_{{\bf q},2}^\dagger,a_{-{\bf q},1}^\dagga,a_{-{\bf q},2}^\dagga\right)$}
and $\bar{H}_{\bf q}^C$ is the $4\times4$ matrix
\begin{equation}
 \bar{H}_{\bf q}^C=
 \left(\begin{array}{cccc}
  \tilde{A}_{\bf q} & \tilde{E}_{\bf q}^\star & \tilde{C}_{\bf q} & \tilde{E}_{\bf q}^\star  \\
  \tilde{E}_{\bf q} & \tilde{B}_{\bf q} & \tilde{E}_{\bf q} & \tilde{D}_{\bf q}\\
  \tilde{C}_{\bf q} & \tilde{E}_{\bf q}^\star & \tilde{A}_{\bf q} & \tilde{E}_{\bf q}^\star \\
  \tilde{E}_{\bf q} & \tilde{D}_{\bf q} & \tilde{E}_{\bf q} & \tilde{B}_{\bf q}
 \end{array}
\right),
\end{equation}
with coefficients
\begin{equation}\label{eq:Coefficients Columnar state}
 \begin{array}{l}
  \tilde{A}_{\bf q}=\displaystyle\frac{1}{2S}\left[2J n_1^z \left(n_2^z-n_1^z\right)+\Gamma n_1^x +J \left(n_1^x\right)^2\cos q_z\right] ,\\
  \tilde{B}_{\bf q}=\displaystyle\frac{1}{2S}\left[2J n_2^z \left(n_2^z+n_1^z\right)+\Gamma n_2^x -J \left(n_2^x\right)^2\cos q_z\right] ,  \\[3mm]
  \tilde{C}_{\bf q}=\displaystyle\frac{J}{2S}\left(n_1^x\right)^2\cos q_z, \quad
  \tilde{D}_{\bf q}=-\displaystyle\frac{J}{2S}\left(n_2^x\right)^2\cos q_z, \\[3mm]
  \tilde{E}_{\bf q}=\displaystyle-\frac{J}{4S}n_1^xn_2^x\left(1+e^{i2q_x}\right). \\
 \end{array}
\end{equation}
Naturally, the classical spin components in Eqs.~(\ref{eq:Coefficients Columnar state}) are those minimizing $E^C({\bf n}_1,{\bf n}_2)$.
For $\Gamma/J\ll 1$ all spins are almost parallel to the $z$ axis, the deviations from it depending on $\Gamma/J$ as
\begin{equation}\label{eq:Columnar state angles low gamma}
\begin{array}{lll}
 n_1^x\approx(\Gamma/J)^{1/3}\ , & & n_2^x\approx \Gamma/4J,
\end{array}
\end{equation}
while just below $\Gamma_c$ all spins are almost parallel to the $x$ axis, the $z$ components behaving as
\begin{equation}\label{eq:Columnar state angles gammac}
\begin{array}{lll}
\displaystyle \left(n_1^z\right)^2\approx\frac{\Gamma_c-\Gamma}{3\left(\sqrt{2}+1\right)J}\ , & & \displaystyle\left(n_2^z\right)^2\approx \frac{\sqrt{2}+1}{3J}(\Gamma_c-\Gamma).
\end{array}
\end{equation}
The term $\tilde{\Delta}_{\bf q}$ in Eq.~(\ref{eq:Fluctuation H columnar}) is defined by $\tilde{\Delta}_{\bf q}=-(\tilde{A}_{\bf q}+\tilde{B}_{\bf q})$.

In the two classical structures that are compared, the fluctuation
Hamiltonians (\ref{eq:Fluctuation H plaquette}) and (\ref{eq:Fluctuation H columnar})
do not contain terms which are linear in the
Holstein-Primakoff bosons and therefore are purely quadratic. In the case
of the plaquette state, the correction to the classical energy is obtained
by diagonalizing the fluctuation Hamiltonian. This is done
via a standard Bogoliubov transformation of (\ref{eq:Fluctuation H plaquette})
which yields the dispersion relations and the correction to the classical energy.

The computation in the case of the columnar state is more involved. This
time the starting classical configuration is not a minimum but a saddle
point of the classical energy. This results in a quadratic fluctuation
Hamiltonian which is not positive definite. An attempt to diagonalize the
quadratic Hamiltonian via a Bogoliubov transformation in this case would yield
a spectrum which is not well defined for some values of momenta. For the
regions in the Brillouin zone near ${\bf q}_B=(0,\pm \pi)$, the spectrum would
not be real but would take non-physical complex values. In such a situation, the correction to the classical energy cannot be computed. The fact that the
spectrum is not well defined results from the truncation of the spin-wave
approximation to the harmonic level. If a state with columnar structure is to become the ground state when quantum fluctuations are fully taken into account, the higher-order terms of the spin-wave expansion must enforce the spectrum to have real frequencies.

In order to avoid the explicit inclusion of the higher-order terms into the analysis, we proceed in the following
way. We add to the Hamiltonian $\mathcal{H}^C$, Eq.~(\ref{eq:Fluctuation H columnar}), the positive term,
\begin{equation} \label{V}
V=\frac{\delta}{S}\sum_i\left[S-S_{i}^{z^\prime(i)}\right]\,,
\end{equation}
describing the presence on each site of an auxiliary field $\delta$ oriented along
the direction of the classical spin at this site, that is, along $z'(i)$.
In Eq.~(\ref{V}), the summation is taken over all sites and $\delta \geq0$ parametrizes the auxiliary field strength.
In terms of Holstein-Primakoff bosons, $V$ takes the form \makebox{$V=(\delta/S)\sum_i a_i^\dagger a_i$}.
The addition of $V$ shifts the coefficients
$\tilde{A}_{\bf q}$ and $\tilde{B}_{\bf q}$ in the columnar-state spin-wave Hamiltonian, Eq.~(\ref{eq:Coefficients Columnar state}), up by the same value,
\begin{equation}
 \begin{array}{l}
\displaystyle \tilde{A}_{\bf q}  \rightarrow \tilde{A}_{\bf q}+\frac{\delta}{2S}, \\[3mm]
\displaystyle \tilde{B}_{\bf q}  \rightarrow \tilde{B}_{\bf q}+\frac{\delta}{2S}, \\[3mm]
 \end{array}
\end{equation}
but does not change any other coefficients.
Accordingly the matrix $\bar{H}_{\bf q}^C$ and the term $\tilde{\Delta}$ become
\begin{equation}
 \begin{array}{l}
\displaystyle \bar{H}_{\bf q}^C \rightarrow \bar{H}_{\bf q}^C+\frac{\delta}{2S}\mathbb{I},\\[3mm]
\displaystyle \tilde{\Delta}_{\bf q} \rightarrow \tilde{\Delta}_{\bf q}-\frac{\delta}{S}. \\
 \end{array}
\end{equation}
Hence the effect of the auxiliary field is to shift all eigenvalues of $\bar{H}_{\bf q}^C$ up by $\delta/2S$.

The value of this auxiliary field is adjusted to obtain a fluctuation Hamiltonian which is non-negatively defined, allowing it to be diagonalized by a Bogoliubov transformation.
This is done by choosing $\delta$ for any given ratio $\Gamma/J$
such that the lowest eigenvalue of $\bar{H}_{\bf q}^C$ is equal to zero.
The resulting spectrum has real and positive frequencies with soft modes only at the wavevector ${\bf q}_B=(0,\pm \pi)$.
The advantages of this approach are the following: the addition of $V$ to the
Hamiltonian $\mathcal{H}^C$ does not change the classical energy of
the state considered [$\mathcal{H}^{(0)}$ is left unchanged] and allows to obtain dispersion relations which are
physically meaningful. Furthermore, $V$ is
a strictly positive contribution to the Hamiltonian, hence the corrections to
the energy of the columnar state computed with this approach provide an {\it
upper} bound for the energy of this state at order $1/S$.

\begin{figure}[htbp]
\centering
 \includegraphics[width=\columnwidth]{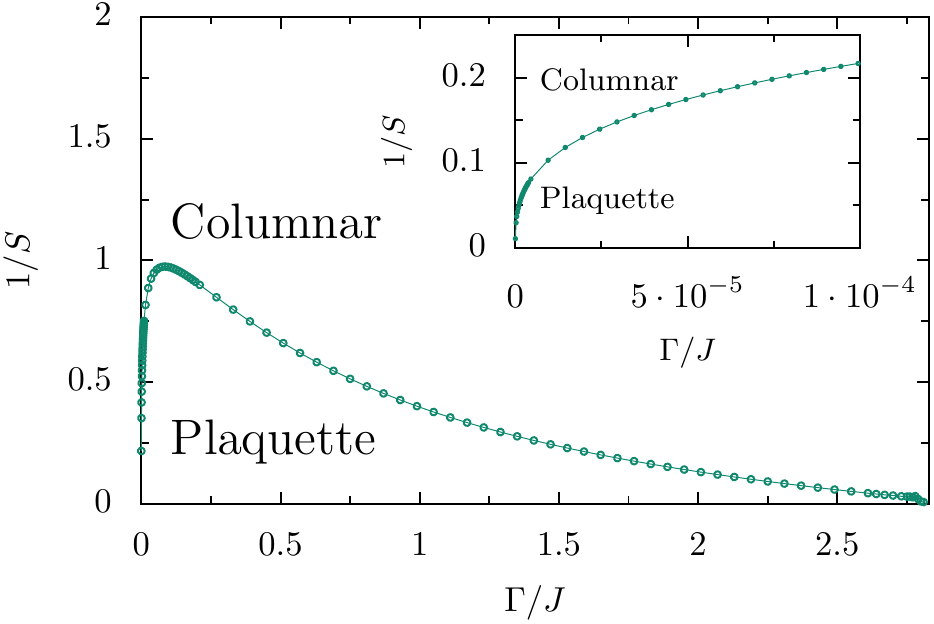}
\caption{(Color online) Phase diagram in the $1/S$ vs. $\Gamma/J$ plane obtained within the framework of the harmonic approach.
The solid line is an upper bound 
of the range of stability of the plaquette state.
In particular, for $S=1/2$ the columnar state is stabilized over the entire range $0\leq\Gamma\leq 2\sqrt{2}J$.
The inset is a zoom of the low field limit.
For any value of $S<\infty$, there exists a finite field range for which the columnar state is stabilized over the
plaquette state.}
\label{fig: Semiclassical Phase Diagram}
\end{figure}

Now we can compare the energy of the plaquette state corrected by the inclusion of the harmonic fluctuations with an estimate from above for the energy of the columnar state calculated to the same order in $1/S$.
We find that as a function of $S$ and $\Gamma/J$, the {\it upper} bound
for the energy of the columnar state is lower than the energy of the
plaquette state in a significant parameter range, see Fig.~\ref{fig: Semiclassical Phase Diagram}.
This approach shows that quantum fluctuations to order $1/S$ easily overcome the energy difference between the
classical energies of the plaquette and columnar structures. This is not surprising since, as already discussed
in Sec.~\ref{subsec:Classical ground states and their dimer representation},
the difference of classical energies of the two states is very small
compared to the energy of the plaquette state.
The auxiliary field $\delta$ required to obtain real and positive frequencies for the columnar state
ranges from $0$ to approximately $0.25J$ and is maximal for $\Gamma/J\approx0.4$.

It has to be noted that at $\Gamma=0$ both columnar and plaquette states are the ground states of the Hamiltonian and remain degenerate for all values of $S$.
The approach developed in this section suggests that for any finite value of $S$ as soon as $\Gamma$ is turned on the harmonic fluctuations favor the columnar
state over the plaquette state.
In fact, at low transverse fields, we find that the critical value of $S$ at which the transition between the two phases occurs scales like $S_c\sim(\Gamma/J)^{1/3}$
(see inset of Fig.~\ref{fig: Semiclassical Phase Diagram}).
Naturally, for small $S$ (including \makebox{$S=1/2$}) the higher-order corrections in $1/S$ may be important. However, when one
disregards them, our analysis predicts that  at $S=1/2$ the columnar state is stabilized for all fields below $\Gamma_c$.

\subsection{Fluctuations and the applicability of the harmonic approximation}\label{Sec:Fluctuations}
In this section we present the analysis of the amplitude of fluctuations and discuss the transverse field domain for which the harmonic approximation is justified.
In fact it is reasonable to wonder whether fluctuations completely melt the classical order of the structures considered. In spin problems, one can distinguish longitudinal
and transverse fluctuations (definitely related to each other). 
By longitudinal fluctuations we refer to the suppression of the spin projection along its average direction.
Since the spin projection along the classical direction is given by $S_i^{z^\prime(i)}=S-a_i^\dagger a_i^\dagga$,
a natural measure of this quantity is given by the average number of Holstein-Primakoff bosons in the harmonic ground state
\begin{equation}
 w=\langle a_i^\dagger a_i^\dagga \rangle\,.
\end{equation}
For a given $S$, longitudinal fluctuations will be small as long as
$w\ll S$. The fulfillment of the same condition ensures that the self consistency criterion necessary to justify the harmonic truncation of the
Holstein-Primakoff transformation (\ref{eq:HP transformation}) is satisfied.

By transverse fluctuations we refer to fluctuations which are transverse to the direction of the classical spins, i.e. the average of $(S^{x^\prime(i)}_i)^2+(S^{y^\prime(i)}_i)^2$ 
where the axes $x^\prime(i)$ and $y^\prime(i)\equiv y$, introduced previously, are orthogonal to $z^\prime(i)$, the direction of the classical spin at site $i$. Comparing the transverse fluctuations to the square of the spin length $S(S+1)$ yields the quantity $\langle (S^{x^\prime(i)}_i)^2+(S^{y^\prime(i)}_i)^2\rangle/S(S+1)$
which is related to $w$ as follows
\begin{equation}\label{eq:TransverseFluctuations HP}
 \frac{\left\langle \left(S^{x^\prime(i)}_i\right)^2+\left(S^{y^\prime(i)}_i\right)^2\right\rangle}{S(S+1)}
=\frac{1+2w}{S+1}+\mathcal{O}\left(\frac{1}{S^2}\right).
\end{equation}
It follows from Eq.~(\ref{eq:TransverseFluctuations HP}) that for $S\gg 1$ the smallness of the transverse fluctuations is ensured by the same condition \makebox{$w\ll S$} as for the longitudinal ones. On the other hand, in contrast to the longitudinal fluctuations, the transverse fluctuations are never fully suppressed: Even in the limit $w\rightarrow 0$ (in which the spins are fully polarized along the $z^\prime$ directions), they remain finite and tend to their minimal value $1/(S+1)$.

\begin{figure}[htbp]
 \centering
\includegraphics[width=\columnwidth]{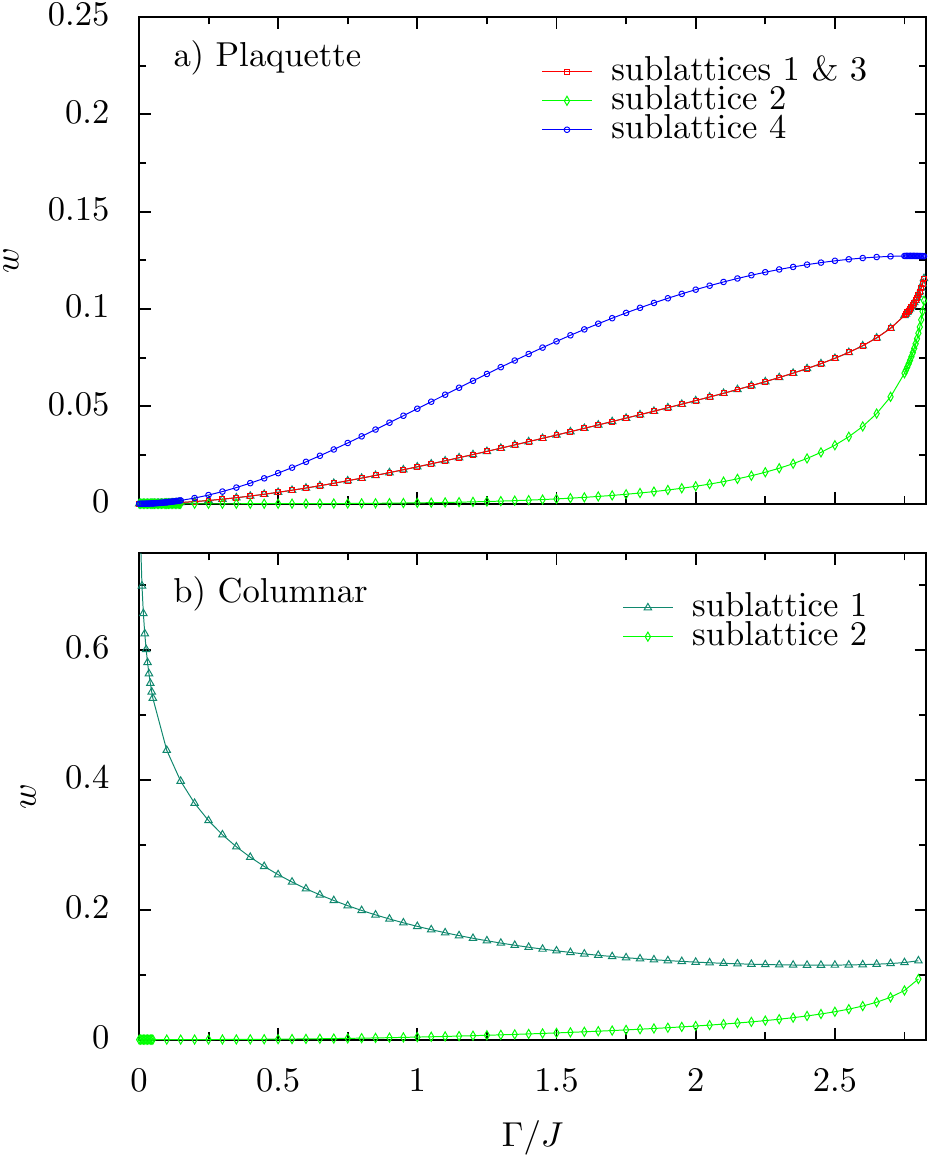}
\caption{(Color online) Plot of the $S$ independent quantity $w=\left\langle a_i^\dagger a_i \right\rangle$
for: a) the plaquette and b) the columnar structures.}
\label{fig:Plot w}
\end{figure}

Since the columnar and plaquette structures considered in our semiclassical calculation consist of several inequivalent sublattices,
we have computed the quantity $w$ for each of them.
The averages $\langle(S^{x^\prime(i)}_i)^2\rangle$ and $\langle(S^{y^\prime(i)}_i)^2\rangle$ are not presented separately because,
almost always (with one exception explicitly mentioned below), the two quantities are comparable to each other.
The plots of $w$ are presented in Fig.~\ref{fig:Plot w}.

For the plaquette state we obtain that $w < 0.13$ for all fields and sublattices. As expected from symmetry, the fluctuations have the same amplitude
on sublattices $1$ and $3$. This is the case since the sites on these two sublattices have the same environment.
In the limit $\Gamma/J\rightarrow0$ (relevant for the mapping to the QDM), the average $\langle a_i^\dagger a_i\rangle$ tends to zero for all sublattices and the transverse fluctuations
are the lowest. This is in agreement with the fact that at $\Gamma=0$ the classical plaquette state is an eigenstate of the Hamiltonian
and the fluctuations, which can be calculated exactly for each spin, are as
low as they can be.
For $\Gamma\ll J$, where the transition between the plaquette and columnar states takes place at $S=S_c(\Gamma/J)\propto(\Gamma/J)^{1/3}$, the value of $w$ at $S=S_c(\Gamma/J)$ satisfies the condition $w\ll S$.
The fulfillment of this criterion means that the spins are almost fully polarized in the corresponding $z^\prime$ directions.
This allows us to expect that the harmonic approximation is applicable for the plaquette state.

In the case of the columnar state the situation is more involved.
The behavior of $w$ for sublattice $2$ is qualitatively similar
to the case of the plaquette state. However, for sublattice $1$, at low fields,
$\langle (S^{x^\prime(i)}_i)^2\rangle$ diverges while $\langle (S^{y^\prime(i)}_i)^2\rangle$
remains finite. This shows up in the low field divergence of $w$ for sublattice $1$, see Fig.~\ref{fig:Plot w}(b).
A priori this does not allow to trust the validity of the approximation that we used for the columnar state
as the classical order is wiped out by the diverging transverse fluctuations.
This divergence is related to the properties of the excitation spectra of the modified Hamiltonian $\mathcal{H}^C+V$, as
explained in detail in Appendix \ref{Sec:AppendixA}.

In view of the divergent fluctuations on sublattice $1$ for the columnar state in the limit $\Gamma/J\rightarrow 0$, it is impossible to expect that the harmonic approximation
is applicable in the small $\Gamma/J$ range. Another manifestation of this comes from observing that in the limit $\Gamma/J\rightarrow 0$ the leading anharmonic contribution
to the energy is also divergent.

A possible way out from the problems related to strong fluctuations could consist in increasing the value of the auxiliary field $\delta$:
so far, this field has been adjusted just to make the spectrum well defined.
Increasing it further will open a gap and remove the divergence of the fluctuations, while increasing the harmonic contribution to the energy at the same time.
By increasing $\delta$ by a few percent, it is indeed possible
to remove the divergence of the fluctuations and make $w < 0.16$ while keeping the harmonic energy low enough to still favor the columnar state over the plaquette state for low values of $S$ including 1/2 (see Appendix \ref{Sec:AppendixB}).

While achieving a finite density of HP particles in the harmonic approximation is a step in the right direction, truncating the expansion to harmonic level is formally justified only if the contribution of anharmonicities is negligible with respect to that of the harmonic terms.
For this purpose we computed the next to leading order correction to the energy for an auxiliary field increased by $5$\textdiscount~above the minimal
value ensuring the harmonic spectrum to be real.
For $\Gamma/J$ not too small ($\Gamma\gtrsim0.6J$), we obtain that the transition between the columnar and plaquette states occurs in the region of the phase diagram
where the $1/S$ expansion is expected to work since the leading anharmonic correction to the energy no longer dominates over the harmonic one (see Appendix \ref{Sec:AppendixB}).
This allows one to conclude that at least in this interval of fields the increase of fluctuations (that is, the decrease in $S$) leads to the phase transitions from the plaquette to columnar state. Accordingly, our results can be considered as an analytical explanation of the numerical results of Ref.~\onlinecite{Wenzel12}.

By contrast, at lower values of $\Gamma$, considering increased auxiliary fields solves only some of the problems but does not lead to a fully consistent theory.
In particular, upon decreasing $\Gamma$, the leading anharmonic contribution to the energy becomes much larger than the harmonic contribution
because it scales with $\Gamma/J$ with an exponent smaller than that of the harmonic contribution (see Appendix \ref{Sec:AppendixB}).
In addition, in the limit $\Gamma/J\rightarrow 0$, the mapping to the square lattice QDM shows that the energy correction induced by the transverse field should be
linear in $\Gamma$, a behavior that it has not been possible to reproduce with adjusting $\delta$ if anharmonic corrections are taken into account.

This suggests that an approach that goes beyond a simple expansion in $1/S$ is necessary to properly treat the limit \makebox{$\Gamma/J\rightarrow 0$},
a task that goes beyond the scope of the present paper. As a consequence, the low $\Gamma/J$ behavior of the transition line between the plaquette and the
columnar phase may be quite different from the $(\Gamma/J)^{1/3}$ prediction following from our analysis. Accordingly, the true form of this
dependence must be considered an open issue.


\section{Conclusion}\label{sec:Conclusion}
To conclude, in this work we have investigated the zero-temperature phase diagram of the fully frustrated transverse field Ising model on the square lattice.
Both analytical arguments and numerical minimizations demonstrate that for all fields $0<\Gamma<\Gamma_c\equiv 2\sqrt{2}J$ the global minimum of the classical energy functional
(describing the system in the $S\rightarrow \infty$ limit) is achieved in the eightfold degenerate configuration having plaquette structure.
At $\Gamma_c$ the system enters the polarized phase with all spins aligned along the field.

For any field in the interval $0<\Gamma<\Gamma_c$, the eight plaquette states belong to a ring of almost degenerate states. Going along this ring, one also passes through eight
configurations with columnar structure which are saddle points of the classical energy. When performing an expansion of the classical energy in the vicinity of $\Gamma_c$
(the expansion up to the eighth order is required to remove the accidental degeneracy), the columnar states emerge as the lowest-energy states if one takes into account only
the critical modes, whereas the plaquette structure is recovered when the interaction of critical modes with subcritical ones is consistently included into the analysis.
The energy difference between the plaquette and columnar solutions in the vicinity of $\Gamma_c$ scales like $|\Gamma-\Gamma_c|^4/J^3$ and is very small in the whole range
$0 <\Gamma <\Gamma_c$.

Zero-point fluctuations around the plaquette and columnar solutions are investigated in the framework of the linear spin-wave approximation.
This is done by introducing a fluctuation Hamiltonian in terms of Holstein-Primakoff bosons which is truncated at the harmonic level.
In the case of the plaquette state the fluctuation Hamiltonian is diagonalized by a Bogoliubov transformation directly yielding the zero-point
corrections to the energy.
The harmonic fluctuation Hamiltonian of the columnar solution, which is a saddle point of the classical energy, is not positive definite, and
proceeding in the same way would lead to imaginary frequencies.
In this case, a positive definite fluctuation Hamiltonian is achieved by the addition of an auxiliary field term which is a positive contribution
to the original Hamiltonian.
The effect of the auxiliary field is to stabilize the columnar state allowing the harmonic fluctuation Hamiltonian to be diagonalized by the standard Bogoliubov approach.
The resulting zero-point energy provides us with an upper estimate for the energy of the state with columnar structure which we compare to the corrected energy of the plaquette state.

Overall we find that the general tendency of harmonic fluctuations is to favor the columnar state over the plaquette state.
At the harmonic level, and in the ultra-quantum limit $S=1/2$ (the value of the spin for which the FFTFIM maps onto the QDM at low fields), the columnar state turns out to be
stabilized in the entire range of fields below saturation, in agreement with the quantum Monte Carlo results of Ref.~\onlinecite{Wenzel12}.
This conclusion is additionally supported by the fact that for not too small $\Gamma/J$ ratios, the phase transition between the columnar and plaquette states occurs at the values of $S$ for which expansion to harmonic order is a reasonable approximation.
Hence we provide further evidence in favor of the transition between the columnar ordered phase and the polarized state.
However, at $\Gamma/J\ll 1$ the harmonic approximation cannot be relied upon and a more sophisticated analytical treatment of this limit is required, which is left for future investigation.
Quite remarkably, the sequence of phase transitions taking place in the spin-1/2 FFTFIM on the square lattice with increasing temperature is expected to be the same independently of whether the ground state has the columnar or plaquette structure.\cite{Korshunov12}

\section*{ACKNOWLEDGMENTS}
We acknowledge useful discussions with Sandro Wenzel. This work has been supported by the Swiss National Foundation.

\appendix
\section{Divergence of fluctuations in the columnar state}\label{Sec:AppendixA}
In this Appendix we explicitly show how the divergence of fluctuations for the columnar state at weak transverse fields follows from the properties of the excitation spectra of the modified Hamiltonian, $\mathcal{H}^C+V$.
By construction, the lowest branch of its excitations has no gap at ${\bf q}_B=(0,\pm\pi)$.
Furthermore, this branch of excitations has a very weak dispersion along the $q_x$ direction.
This fact, together with the vanishing of the gap, results in the presence in the Brillouin zone of a line of very low-energy excitations.
This can be seen analytically by considering the quadratic fluctuation Hamiltonian (\ref{eq:Fluctuation H columnar}) in the low-field limit.
Using the low-field expressions of the classical spin components in the columnar state, Eq. (\ref{eq:Columnar state angles low gamma}),
we obtain that to the order $(\Gamma/J)^{2/3}$ the coefficients $\tilde{A}_{\bf q}$, $\tilde{B}_{\bf q}$ and $\tilde{C}_{\bf q}$ in Eq.~(\ref{eq:Coefficients Columnar state}) are given by
\begin{equation}
\begin{array}{l}
\displaystyle \tilde{A}_{\bf q}\approx J\frac{(\Gamma/J)^{2/3}}{2S}(1+\cos{q_z})\,, \\ [2mm]
\displaystyle \tilde{B}_{\bf q}\approx\frac{J}{2S}\left[4-(\Gamma/J)^{2/3}\right]\,, \\ [2mm]
\displaystyle \tilde{C}_{\bf q}\approx J\frac{(\Gamma/J)^{2/3}}{2S}\cos{q_z}\,,
\end{array}
\end{equation}
whereas the coefficients $\tilde{D}_{\bf q}$ and $\tilde{E}_{\bf q}$ can be neglected. Within this approximation, the harmonic fluctuations on the two sublattices are now decoupled from each other,
\begin{equation}\label{eq:H col in low field}
\begin{array}{ll}
 \vec{a}_{\bf q}^\dagger \bar{H}_{\bf q}^C\vec{a}_{\bf q} \approx &\tilde{A}_{\bf q}(a_{{\bf q},1}^\dagger a_{{\bf q},1}+a_{{-\bf q},1} a_{{-\bf q},1}^\dagger)
+\tilde{C}_{\bf q}(a_{{-\bf q},1} a_{{\bf q},1}+\textrm{h.c.}) \\
&+ \tilde{B}_{\bf q}(a_{{\bf q},2}^\dagger a_{{\bf q},2}+a_{{-\bf q},2} a_{{-\bf q},2}^\dagger)\,,
\end{array}
\end{equation}
with the eigenvalues of $\bar{H}_{\bf q}^C$ being $\tilde{B}_{\bf q}$ (twice degenerate) and \makebox{$\tilde{A}_{\bf q}\pm \tilde{C}_{\bf q}$}. While $\tilde{B}_{\bf q}$ and $\tilde{A}_{\bf q}-\tilde{C}_{\bf q}$
are always positive, the eigenvalue
\begin{equation}\label{eq:Low gamma non positive eigenvalue}
 \tilde{A}_{\bf q}+ \tilde{C}_{\bf q}\approx J\frac{(\Gamma/J)^{2/3}}{2S}(1+2\cos{q_z})
\end{equation}
is negative in a substantial part of the Brillouin zone. The minimal auxiliary field amplitude required to achieve a positive definite quadratic form is $\delta\approx J(\Gamma/J)^{2/3}$. As a result, the diagonal elements of $\bar{H}_{\bf q}^C$
get a positive correction $\delta/2S$,
 \begin{equation}
 \begin{array}{l}
  \displaystyle \tilde{A}_{\bf q}\rightarrow\tilde{A}_{\bf q}
 +\frac{\delta}{2S}\approx J\frac{(\Gamma/J)^{2/3}}{2S}(2+\cos{q_z})\,,
 \\ [2mm]
 \displaystyle \tilde{B}_{\bf q}\rightarrow\tilde{B}_{\bf q}
 +\frac{\delta}{2S}\approx\frac{2J}{S}\,. \\ [2mm]
 \end{array}
 \end{equation}

Furthermore, in this approximation the part of the Hamiltonian (\ref{eq:H col in low field}) involving Holstein-Primakoff bosonic operators on sublattice $2$ is already diagonal
in these operators.
This explains why in the ground state the value of $\langle a_i^\dagger a_i\rangle$ on sublattice $2$ sites tends to zero at low fields and accordingly $w$ tends to $0$.
The remaining part of Eq.~(\ref{eq:H col in low field}), involving  bosonic operators on sublattice $1$, can be diagonalized by a standard Bogoliubov transformation.
It is then straightforward to obtain the average number of Holstein-Primakoff bosons on sublattice $1$ sites ($w_1\equiv \langle a_{i,1}^\dagger a_{i,1} \rangle$)
\begin{equation}\label{eq:Fluctuations col state}
w_1 \approx \frac{1}{2N}\sum_{\bf q}\frac{\tilde{A}_{\bf q}+\delta/2S}{\sqrt{(\tilde{A}_{\bf q}+\delta/2S)^2-\tilde{C}_{\bf q}^2}}-1.
\end{equation}
From these low-field considerations we observe that the summand in the right-hand side of  Eq.~(\ref{eq:Fluctuations col state})
has no dispersion along the $q_x$ direction at this level of approximation.
Furthermore,
the term $\tilde{A}_{\bf q}+\delta/2S+\tilde{C}_{\bf q}\approx J(\Gamma/J)^{\frac{2}{3}}(1+\cos{q_z})/S$
in the expression standing under the square root
causes the denominator to behave as $|q_z\mp\pi|$ in the vicinity of $q_z=\pm\pi$.
This fact in conjunction with the absence of dispersion along $q_x$ leads to a logarithmically divergent integral. In a more accurate calculation this divergence is
cut by the presence of higher-order corrections, but when $\Gamma/J$ approaches zero it is recovered and the number of Holstein-Primakoff particles on sublattice 1 sites
has to diverge.

The fact that $\langle a_i^\dagger a_i\rangle$ does not tend to zero on sublattice 1 sites when $\Gamma\rightarrow0$
might be surprising at first sight since the columnar state is a ground state of the model at $\Gamma=0$. However, this contradiction can be explained by
realizing that as soon as the field $\Gamma$ is turned on, the bosonic fluctuation Hamiltonian involving sublattice 1 operators becomes essentially nondiagonal since
both coefficients, $\tilde{A}_{\bf q}$ and $\tilde{C}_{\bf q}$, scale in the same way as $(\Gamma/J)^{2/3}$.
So there is no characteristic transverse field scale for which $\tilde{C}_{\bf q}$ is negligible with respect to $\tilde{A}_{\bf q}$.

\section{Assessing the validity of the harmonic approximation for the columnar state}\label{Sec:AppendixB}
The study of fluctuations in Sec.~\ref{Sec:Fluctuations} signals the breakdown of the harmonic approximation for the columnar state at low transverse fields.
In the limit $\Gamma/J\rightarrow0$, the average number of Holstein-Primakoff bosons on sublattice $1$, $w_1$, diverges
[see Fig.~\ref{fig:Plot w}b)], as well as
the leading anharmonic contribution to the energy of this state.

In order to cut the divergence of $w_1$, we attempted to increase the value of the auxiliary field $\delta$.
In the results presented so far, $\delta$ was always tuned to the value $\delta_0(\Gamma/J)$ defined as the minimal auxiliary field amplitude
which makes the spin-wave spectrum well defined but gapless at a given ratio $\Gamma/J$.
As already mentioned in Sec.~\ref{Sec:Fluctuations}, an increase of the auxiliary field strength above its minimal value will open a gap in the spectrum and cut the divergence leading to a decrease of fluctuations.
If one increases the auxiliary field from $\delta_0(\Gamma/J)$ to
\begin{equation}\label{eq:Local Field increment}
 \delta(\Gamma/J)= (1+\epsilon)\delta_0(\Gamma/J)
\end{equation}
with $\epsilon>0$, the sum in Eq.~(\ref{eq:Fluctuations col state}) is no longer divergent, which allows one to write down an explicit expression for
the average number of Holstein-Primakoff bosons on sublattice $1$ in the limit $\Gamma/J\rightarrow0$,
\begin{equation}\label{eq:Fluctuations col state CUTOFF}
 w_1\approx \frac{1}{2N}\sum_{\bf q}\frac{2+\epsilon +\cos q_z}{\sqrt{(2+\epsilon)(2+\epsilon+2\cos q_z)}}-1.
\end{equation}

\begin{figure}[htbp]
 \centering
\includegraphics[width=\columnwidth]{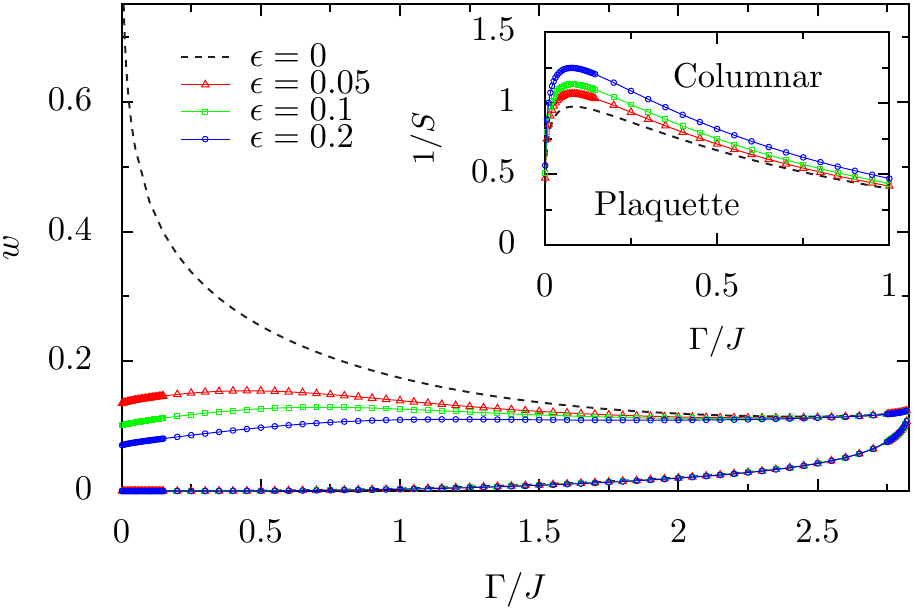}
\caption{(Color online) Plot of the $S$ independent quantity $w$ for the columnar state for different auxiliary fields.
Triangles, squares and circles respectively correspond to auxiliary field increments of $5,10$ and $20$ percent.
The dashed lines correspond to results obtained for the minimal auxiliary field amplitude required to obtain a real spectrum of
excitations for the columnar state. (Inset) Plot of the estimate from above for the boundary between the plaquette and columnar structures
in the $1/S - \Gamma/J$ plane for various auxiliary fields.}
\label{fig:Fluctuations New Field}
\end{figure}

The average numbers of Holstein-Primakoff bosons on both sublattices computed for different values of $\epsilon$ in Eq. (\ref{eq:Local Field increment})
are plotted in Fig.~\ref{fig:Fluctuations New Field}.
We considered auxiliary field increments of $5$, $10$ and $20$ percent.
An increment of $5${\textdiscount} is already sufficient to decrease the fluctuations enough to make $w_1\approx0.125$ at $\Gamma=0$.
As expected, the effect of the increase of the auxiliary field on the fluctuations for sublattice $2$ sites is negligible for low $\Gamma$.

Apart from decreasing the fluctuations, an increase of the auxiliary field penalizes the columnar state by increasing its zero point energy.
Accordingly, the line in the $1/S$ vs. $\Gamma/J$ plane designating the upper boundary of the region of stability of the plaquette state is shifted
to larger values of $1/S$.
However we verified that, for the auxiliary field increments discussed above, this shift does not modify qualitatively the phase diagram.
In the harmonic approximation, the columnar state still has the lowest energy over the entire field range when $S=1/2$, as can be seen in the
inset of Fig.~(\ref{fig:Fluctuations New Field}).

Nevertheless, achieving a finite density of HP particles in the harmonic approximation is not sufficient to conclude that the
semiclassical calculation predicts columnar order. Truncating the expansion to harmonic level is formally justified only
if the contribution of anharmonicities is negligible with respect to that of the harmonic terms.

Keeping in the expansion terms which are next to leading order in the Holstein-Primakoff transformation (\ref{eq:HP transformation})
produces the cubic and quartic bosonic contribution to the Hamiltonian (which we denote respectively $\mathcal{H}^{(3)}\propto S^{-3/2}$ and $\mathcal{H}^{(4)}\propto S^{-2}$
following the notation introduced above).
The next to leading order contribution to the energy, which is of order $1/S^2$, can be obtained by treating $\mathcal{H}^{(3)}$ and $\mathcal{H}^{(4)}$
as perturbations to the harmonic Hamiltonian.

The leading contribution to the energy from
$\mathcal{H}^{(4)}$ consists of its average in the harmonic ground state.
This quantity is readily computed by considering all possible decouplings of the four-body operators.
On the contrary, $\mathcal{H}^{(3)}$, which contains only three-particle terms, contributes to the energy only in the second order of the perturbation theory.
In the absence of degeneracies, the overall $1/S^2$ contribution to the energy is given by
\begin{equation}\label{eq:Next to leading order energy}
 \langle0|\mathcal{H}^{(4)}|0\rangle-\sum_{|e\rangle}\frac{|\langle e|\mathcal{H}^{(3)}|0\rangle|^2}{E_{|e\rangle}-E_{|0\rangle}}
\end{equation}
where $|0\rangle$ and $|e\rangle$ denote respectively the ground state of the harmonic Hamiltonian and excited states.
$E_{|0\rangle}$ and $E_{|e\rangle}$ are the harmonic energies of the ground state and of the excited states.

\begin{figure}[htbp]
 \centering
\includegraphics[width=0.9\columnwidth]{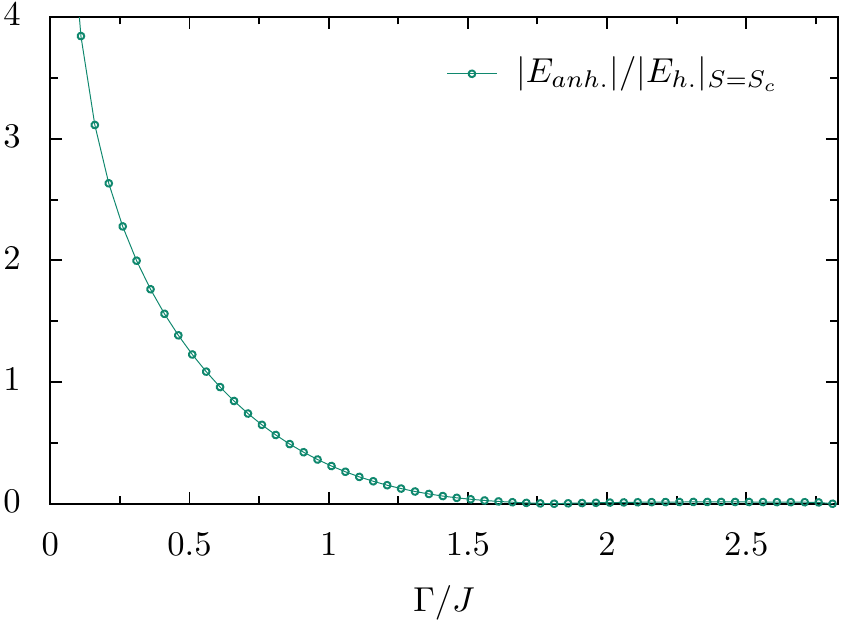}
\caption{(Color online) The ratio of the $1/S^2$ contribution to the columnar state energy to the $1/S$ contribution calculated at $S=S_c(\Gamma/J)$, where the energy of the plaquette state (in the harmonic approximation) is equal to our boundary from above for the energy of the columnar state.
Both energies are computed for the auxiliary field $\delta(\Gamma/J)$ which exceeds by $5$\textdiscount~ the minimal value required to ensure a meaningful spectrum.}
\label{fig:Higher order energy}
\end{figure}

We computed numerically the $1/S^2$ contribution to the energy (\ref{eq:Next to leading order energy}) for a $5$\textdiscount~increment of the auxiliary field above the minimal value
[i.e., with auxiliary field $\delta(\Gamma/J)$ given by Eq.~(\ref{eq:Local Field increment}) with $\epsilon=0.05$].
Figure (\ref{fig:Higher order energy}) is a plot of the $1/S^2$ contribution to the energy divided by the zero point energy correction.
The ratio is evaluated at $S=S_c(\Gamma/J)$, that is, on the line which gives an upper boundary for the region of stability of the plaquette state in the framework of the harmonic approximation.
As can be seen in the figure, this ratio exceeds the value $1$ for transverse fields $0<\Gamma/J\lesssim0.6$.

For larger values of the transverse field the anharmonic correction to the energy is smaller than the harmonic contribution
at $S=S_c(\Gamma/J)$, thus the transition between the columnar and plaquette states occurs in a region of the phase diagram where the harmonic analysis is applicable.
However, for $S$ essentially below $S_c(\Gamma/J)$, this will no longer be the case consistently with the quantum Monte Carlo results
of Ref.~\onlinecite{Wenzel12} which indicate a suppression by almost a factor 2 of the critical field at which the transition into the polarized state occurs for $S=1/2$ in comparison with the classical transition.
However, we see no reasons to expect that the decrease in $S$ may induce a reentrant transition to the plaquette phase.

At lower fields, we no longer can be sure that the line separating the columnar and plaquette phases belongs to the region where the perturbation theory works,
which makes it impossible to make as definite statements as for higher fields.
In this case we only observe that the first quantum corrections show the tendency for the stabilization of the columnar state, which is compatible with
the results of the Monte Carlo study of the $S=1/2$ problem. \cite{Wenzel12}

\bibliography{bibliography}

\end{document}